\documentclass[aps,superscriptaddress,longbibliography]{revtex4-2}
\usepackage{amsmath}
\usepackage{graphicx,textcomp}
\usepackage{enumerate,subeqnarray}
\usepackage{verbatim,natbib}
\usepackage{amsmath, amsthm, amssymb}
\usepackage{amsfonts}
\usepackage{caption}
\usepackage[colorlinks=true, citecolor=red, linkcolor=blue, urlcolor=blue ]{hyperref}

\newcommand{\oo}[1]{\mathcal O}

\begin{document}
	
	\title{Cortex-driven cytoplasmic flows in elongated cells: fluid mechanics and application to nuclear transport in {\it Drosophila} embryos}
	
	\author{Pyae Hein Htet}
	\affiliation{Department of Applied Mathematics and Theoretical Physics, University of Cambridge,	
		Cambridge CB3 0WA, UK}
	\author{Eric Lauga}
	\email{e.lauga@damtp.cam.ac.uk}
	\affiliation{Department of Applied Mathematics and Theoretical Physics, University of Cambridge,	
		Cambridge CB3 0WA, UK}
	
	\date{\today}
	
	\begin{abstract}
		
		The \textit{Drosophila melanogaster} embryo, an elongated multi-nucleated cell, is a classical model system for eukaryotic development and morphogenesis.  Recent work has shown that bulk cytoplasmic flows, driven by cortical contractions along the walls of the embryo, enable the uniform spreading of nuclei along the anterior-posterior (AP) axis necessary for proper embryonic development. Here we  propose  two mathematical models to characterise cytoplasmic flows driven by tangential cortical contractions in  elongated cells.  Assuming Newtonian fluid flow at low Reynolds number in a spheroidal cell, we first compute the flow field exactly, thereby bypassing the need for numerical computations. We then  apply our results to   recent experiments on nuclear transport in cell cycles 4-6 of \textit{Drosophila} embryo development. By fitting the cortical contractions in our model to  measurements,  we  reveal that experimental cortical flows enable near-optimal  axial spreading of nuclei. A second mathematical approach, applicable to general elongated cell geometries, exploits a long-wavelength approximation to produce an even simpler solution,  with errors below  $5\%$ compared to the full model. An application of this long-wavelength result to  transport leads to fully  analytical solutions for the nuclear concentration that capture the essential physics of the  system, including optimal axial spreading of nuclei.

	\end{abstract}

	\maketitle
	%%%%%%%%%%%
	\section{Introduction}

	The cell cortex, present in most animal cells, is a thin network of polymeric filaments  (actin), molecular motors (myosin), and filament-binding proteins, located directly underneath the cell membrane separating the interior of biological cells from the outside environment~\cite{salbreux2012,chugh2018}. The actin polymer network gives the cell its shape and stiffness, while  myosin motors exert contractile stresses on actin filaments, enabling the active cortical contractions crucial for the cell's mechanical and morphological functions~\cite{salbreux2012,chugh2018}.   
	The control of cell mechanics underlies numerous biological processes. This includes cell migration via cortical flow-generated propulsive forces~\cite{chabaud2015, bergert2015, ruprecht2015, liu2015}, often aided by the formation of forward protrusions called blebs~\cite{paluch2013, blaser2006}. Similarly, the cortex  plays a key role in the shape changes involved in cell division such as mitotic rounding~\cite{stewart2011} and furrow formation and constriction~\cite{schwayer2016}.
	
	An important class of problems in early developmental biology concerns intracellular flows driven by cortical movements~\cite{lu2022}.  	 Much of what is known about these cytoplasmic flows   (the cytoplasm is the complex fluid filling the inside of biological cells)  has been obtained from work on model organisms, including the   fruit fly \textit{Drosophila melanogaster}, the nematode \textit{Caenorhabditis elegans} and the zebrafish {\it Danio rerio}. {This type of   system belongs to a broader class of ``cytoplasmic streaming" problems, where large-scale flows  are induced in large eukaryotic cells, including algae, plants, amoeba, fungi, and -- of particular interest in this paper --  animal cells in early development \cite{goldstein2015,illukkumbura2020,pieuchot2015}.}

	Waves of actin polymerisation along the cortex of the zebrafish zygote (i.e.~fertilized egg) create the flows responsible for cytoplasmic transport of yolk granules~\cite{fuentes2010,shamipour2019}. In \textit{C.~elegans} zygotes, actin-myosin contractions play a crucial role in cell polarisation (i.e.~the creation of asymmetry in cellular organisation)~\cite{gubieda2020}. The cortex flows towards the anterior pole and thus creates cytoplasmic flows along the cell axis directed towards the posterior end, distributing cortical and cytoplasmic cellular components asymmetrically in the anterior-posterior (AP) direction.

\begin{figure}[t!]
		\includegraphics[width=0.6\textwidth]{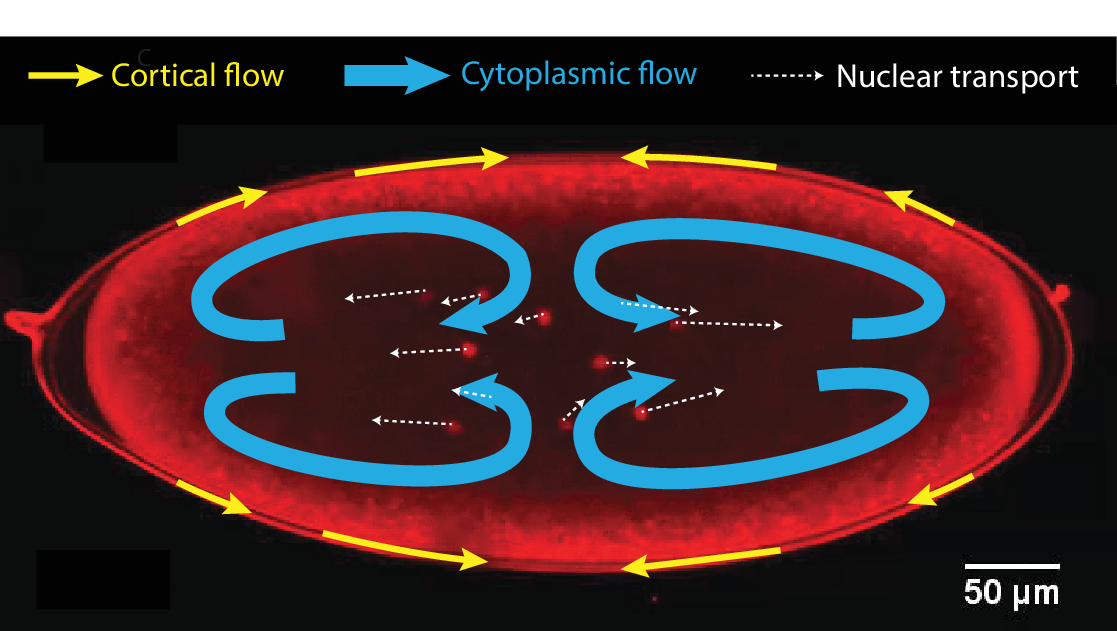}
		\caption{An experimental image of cytoplasmic transport in  \textit{Drosophila} embryo, reprinted from \textit{Cell}, \textbf{177}, Deneke \textit{et al.}, ``Self-Organized Nuclear Positioning Synchronizes the Cell Cycle in \textit{Drosophila} Embryos", 925, Copyright (2019), with permission from Elsevier (Supplementary video 3) \cite{deneke2019}. Red dots located near the centre are nuclei produced by cell division. Arrows are overlaid onto the image to schematically indicate   cortical flows at the edge of the cell (thin yellow arrows),  cytoplasmic flow inside the embryo (thick blue arrows), and the resultant transport of nuclei (thin dotted arrows).}\label{fig:intro}
	\end{figure}

	In this paper, we consider the specific case of \textit{Drosophila} embryos, which are ``syncytial'', meaning that the numerous nuclei obtained in the early few cycles of cell division are not separated into individual cells but instead  share the same cytoplasm. In early development of the embryo, the cortex contracts towards the centre of the cell, most prominently in cell cycles 4-6 (cell cycle $n$ refers to the $n^{\text{th}}$ cycle of cell division), thereby driving cytoplasmic flows that spread the daughter nuclei along the anterior-posterior axis of the embryo, ensuring a uniform positioning of nuclei across the embryo~\cite{deneke2019}. This process is illustrated in  Fig.~\ref{fig:intro} (reproduced  from Supplementary Video 3 in Ref.~\cite{deneke2019}) showing the cortical flows at the edge of the cell (thin yellow arrows), and the resulting cytoplasmic flow inside the embryo (thick blue arrows), which then transports the nuclei generated by cell division (thin dotted arrows). 
	
	The rich mechanobiology of the cell cortex has attracted interest from the physics and mathematical community, spurring the development of active gel theories in recent years, which describe the cortex as a thin layer of viscoelastic material with active tension~\cite{prost2015,salbreux2017,rocha2022,torressanchez2019,bacher2021}. 	Such a framework may be coupled with a model for the flow to reproduce the full dynamics of the surrounding fluid. Recent work couples  active gel theory with a computational  method for the flow (the immersed boundary method) to investigate how a cell deforms in an external flow and  simulate cytoplasmic flows generated by cortical movement~\cite{bacher2019}.

	While these methods are able to model cortical dynamics in great detail, solving the active gel equations is often a complicated and computationally intensive process~\cite{bacher2021}. 	The real biological system is very complex, consisting of an active cortical actin-myosin mesh coupled mechanically and biochemically to the cytoplasm and its components, and modelling all these details faithfully is inevitably a  task that  requires a computational approach~\cite{lopez2023}. 
	
	Biophysical insight could, however, be obtained from simplified models amenable to analytical treatment. In particular, despite its complex chemical composition, the cytoplasm can be described as  an effective Newtonian fluid over timescales of interest, a modelling assumption  supported by past works comparing flow computations with  PIV data~\cite{klughammer2018, deneke2019, niwayama2011} and by rheological measurements \cite{ganguly2012}, {while the forcing from the cortex may be modelled as an effective boundary slip velocity \cite{niwayama2011, deneke2019}. Naturally, such a simple approach has its limitations, and is not able to explain all quantitative features of the cytoplasmic flow.
		For instance, the locations of the vorticity extrema  in the experimentally measured flow in  the \textit{Drosophila} embryo and the required cortical flow velocities deviate from the predictions from a simple Stokes flow (i.e.~a Newtonian fluid flow in the absence of inertia), and a more detailed two-fluid model has been proposed to rectify these disparities \cite{lopez2023}. Nonetheless, a Stokes flow model does reproduce well the large-scale features of the flow %and the important physical picture 
		\cite{deneke2019} and its mathematical simplicity enables further analytical development.}
	
	{For axisymmetric incompressible flow in a spherical geometry, the Stokes equations may} be solved exactly using a series expansion of the streamfunction~\cite{happel1965}.  This general solution has been used to study the axisymmetric flow induced inside a spherical shell by small radial deformations of the boundary, as a model for the cytoplasmic flows generated in  starfish oocytes by cortical contractions~\cite{klughammer2018}.

	However, often the relevant biological cells are elongated, as in the case of \textit{Drosophila} embryos and \textit{C.~elegans} zygotes, introducing an additional level of geometrical complexity to simplified spherical models. Motivated by the problem of cortical contractions in  syncytial \textit{Drosophila} embryos, and the resulting flow-based  transport of nuclei~\cite{deneke2019}, we propose in this paper two analytical mathematical methods to model the cytoplasmic flows generated inside an elongated cell created by arbitrary axisymmetric cortical flow.
	
	Our first approach consists of solving exactly for the {Stokes flow} in a 	spheroidal model cell driven at the boundary {via a slip velocity}. We then show how this three-dimensional flow solution can be applied to transport in  \textit{Drosophila} embryo  to  obtain new insight on intracellular transport. {The second method, valid for any elongated shape, uses a long-wavelength approximation to produce an even simpler, yet remarkably accurate, lubrication solution for the cytoplasmic flows. We exploit this result} to construct, and analytically solve, a reduced model of nuclear transport capturing all the main features of nuclear spreading along the embryo axis.

	Our paper is organised as follows. We first construct the simplest possible Stokes flow model  able to capture the relevant physics of the problem. We obtain analytical solutions for the bulk flow resulting from cortical contractions  on the cell boundary without the need for a numerical approach  (\S\ref{section:maths}). We  then use our solution to reveal that the cortical flows in the \textit{Drosophila} embryo are finely tuned to ensure the optimal spreading of nuclei along the anterior-posterior axis (\S\ref{section:application}). We next derive a long-wavelength solution applicable to any elongated shape (\S\ref{section:lubrication}). Finally, we  use the long-wavelength flow to analytically solve a one-dimensional transport model which reproduces all important features of the full mathematical model (\S\ref{section:1D}).

	%%%%%%%%%%%
	\section{Model for cytoplasmic streaming in elongated cells: Boundary-driven Stokes flows in a prolate spheroid}\label{section:maths}
	
	In order to capture cytoplasmic streaming in elongated cells, we propose a first mathematical model where the cell is represented by a prolate spheroid and the flow, assumed to be Newtonian and in the Stokes regime, is driven from the boundary. 	After detailing the specific problem statement (\S\ref{section:statement}), we introduce prolate spheroidal coordinates  (\S\ref{section:coords}) and the streamfunction in these coordinates (\S\ref{section:streamfunction}). We solve for the flow analytically using a series solution for the streamfunction  (\S\ref{section:solution}) and    illustrate the first few modes of the resulting flow field in terms of the appropriate basis function decomposition of the prescribed tangential slip velocities (\S\ref{section:modes}). 
	
	\begin{figure}[t]
		\includegraphics[width=0.25\textwidth]{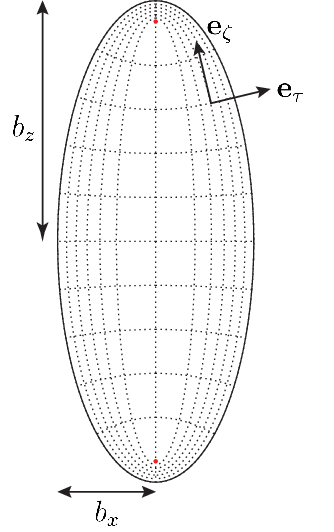}
		\caption{Cross-section containing the long axis of a prolate spheroid of semi-major axis $b_z$ and semi-minor axis $b_x$, illustrating the modified prolate spheroidal coordinate system $(\tau,\zeta,\phi)$ and the basis vectors $\mathbf e_\tau$ and $\mathbf e_\zeta$. Since the flow is axisymmetric, the azimuthal angle $\phi$  is omitted from illustration. Dotted lines are isosurfaces of $\tau$ and $\zeta$. The red dots are the foci at $z = \pm c$.}\label{coord}
	\end{figure}
	
	\subsection{Problem statement}\label{section:statement}
	We consider an incompressible Newtonian fluid, of dynamic viscosity $\mu$, inside a rigid prolate spheroidal domain oriented along the $z$-axis, of semi-major axis $b_z$ and semi-minor axis $b_x < b_z$ (see sketch in Fig.~\ref{coord}). The flow field $\mathbf u$ inside the spheroid is driven by a prescribed axisymmetric tangential slip velocity $\mathbf v_s$ along the boundary. We assume that the relevant velocity scale $U$ and length scale $L$ are sufficiently small such that we may neglect inertia, i.e.~the Reynolds number $Re = \rho UL/\mu \ll 1$ (where $\rho$ is the mass density of the fluid). The fluid flow is then governed by the incompressible Stokes equations
	\begin{equation}
		\mu\nabla^2\mathbf u = \nabla p,\quad\nabla \cdot \mathbf u = 0,
	\end{equation}
	where $p$ is the pressure field, subject to the no-slip  boundary condition
	\begin{equation}\label{eq:no-slip}
		\mathbf u = \mathbf v_s
	\end{equation}
	on the surface of the spheroid.
	
	\subsection{Coordinate system}\label{section:coords}

	We introduce a prolate spheroidal coordinate system~\cite{happel1965} $(\eta,\theta,\phi)$ with semi-focal distance $c := \sqrt{b_z^2 - b_x^2}$, related to the cylindrical polar coordinates $(r,z,\phi)$ via
	\begin{align}\nonumber
		r &= c \sinh \eta \sin \theta, \\
		z & = c \cosh \eta \cos \theta,
	\end{align}	
	and further make the transformation
	\begin{equation}
		\tau = \cosh \eta, \quad \zeta = \cos \theta.
	\end{equation}
	In these modified prolate spheroidal coordinates~\cite{happel1965}, the boundary of the domain is given by 
	\begin{equation}
		\tau = \tau_0 := \frac{b_z}{c},
	\end{equation}
	and the interior of the domain corresponds to $1 \leq \tau < \tau_0$~\cite{pohnl2020}. Surfaces of constant $\tau$ are confocal prolate spheroids, with the degenerate spheroid $\tau = 1$ being the line segment connecting the foci $z = \pm c$. The basis vectors $(\mathbf e_{\tau},\mathbf e_{\zeta})$ in these modified prolate spheroidal coordinates are related to cylindrical polar coordinates by
	\begin{align}
		\mathbf e_\tau &= \frac{\sqrt{\tau^2 - 1}}{\sqrt{\tau^2 - \zeta^2}}\left(\frac{\tau\sqrt{1 - \zeta^2}}{\sqrt{\tau^2 - 1}}\mathbf e_r + \zeta\mathbf e_z\right),\\
		\mathbf e_\zeta &= \frac{\sqrt{1 - \zeta^2}}{\sqrt{\tau^2 - \zeta^2}}\left(\frac{-\zeta\sqrt{\tau^2 - 1}}{\sqrt{1 - \zeta^2}}\mathbf e_r + \tau\mathbf e_z\right).
	\end{align}
	
	The Lam\'e metric coefficients $h_{\phi},h_\zeta,$ and $h_{\tau}$ associated with the prolate spheroidal coordinates are given by
	\begin{align}
		h_\phi &= c\sqrt{\tau^2 - 1}\sqrt{1 - \zeta^2}, \\
		h_\zeta &= c\frac{\sqrt{\tau^2 - \zeta^2}}{\sqrt{1 - \zeta^2}},\\
		h_\tau &= c\frac{\sqrt{\tau^2 - \zeta^2}}{\sqrt{\tau^2 - 1}}.
	\end{align}
	
	The setup and  coordinate system are illustrated in Fig.~\ref{coord}. 
	
	\subsection{Streamfunction}\label{section:streamfunction}
	Exploiting the axisymmetry of the problem, we seek a solution in terms of a Stokes streamfunction $\psi(\tau,\zeta)$ satisfying $\mathbf u = \nabla \times \left({\psi}\mathbf e_\phi / {h_\phi}\right)$~\cite{happel1965}, or more explicitly,
	\begin{equation}\label{u}
		\mathbf u = u_\tau \mathbf e_\tau + u_\zeta \mathbf e_\zeta = \frac{1}{h_\zeta h_\phi} \frac{\partial \psi}{\partial \zeta}\mathbf e_\tau -\frac{1}{h_\tau h_\phi} \frac{\partial \psi}{\partial \tau} \mathbf e_\zeta.
	\end{equation}
	
	Note that a streamfunction formulation automatically ensures incompressibility, $\nabla \cdot \mathbf u = 0$. 
	
	{Writing the tangential slip velocity at the boundary as $\mathbf v_s = v_s\mathbf e_\zeta$, we have two boundary conditions for the streamfunction: (i) impenetrability,
		\begin{equation}
			u_\tau(\tau_0,\zeta) = 0,
		\end{equation}
		and (ii) the prescribed slip velocity}, 
	\begin{equation}
		u_\zeta(\tau_0,\zeta) = v_s(\zeta).
	\end{equation}
	
	\subsection{Solution}\label{section:solution}
	The general `semiseparable' solution for the streamfunction was derived by \citet{dassios1994}. Recently, P\"ohnl \textit{et al.} \cite{pohnl2020} used this general solution to solve for the swimming velocity of and flow field outside a spheroidal squirmer with an axisymmetric tangential slip velocity. In the first mathematical part of the current paper,  we adapt this approach to solve the complementary `interior' problem.

	The general solution for the streamfunction is given by
	\begin{equation}\label{eq:general}
		\psi(\tau,\zeta) = g_0(\tau)G_0(\zeta) + g_1(\tau)G_1(\zeta) + \sum_{n = 2}^{\infty} [g_n(\tau)G_n(\zeta) + h_n(\tau)H_n(\zeta)],
	\end{equation}  
	where $G_n$ and $H_n$ are Gegenbauer functions of the first and second kind respectively~\cite{abramowitz2013}, and $g_n$ and $h_n$ are specific linear combinations of $G_n$ and $H_n$.

	Now the $H_n$'s are singular at $\zeta = \pm 1$ (i.e.~along the $z$-axis, between the poles and the foci), thus requiring $h_n = 0$ for all $n$. $G_0(\zeta) = 1$ and $G_1(\zeta) = -\zeta$ are non-zero at $\zeta = \pm 1$ so we also require $g_0 = g_1 = 0$; otherwise $v_\zeta$ would diverge at $\zeta = \pm 1$  because of the Lam\'e metric coefficients. Non-zero $g_n$'s are allowed for $n \geq 2$, however, since $G_{n\geq2}(\pm1) = 0$. The general solution from Eq.~\eqref{eq:general} therefore reduces to 
	\begin{equation}\label{psi}
		\psi(\tau,\zeta) = \sum_{n \geq 2} g_n(\tau)G_n(\zeta).
	\end{equation}
	Upon further using a similar condition that the terms $G_0(\tau)$ and $G_1(\tau)$ terms cannot be present since the solution would otherwise diverge at $\tau = 1$, the admissible terms in the $g_n$'s are 
	\begin{subequations}\label{g}
		\begin{align}
			\label{g2}
			g_2(\tau) &=  F_2G_2(\tau) + E_4G_4(\tau), \\
			g_3(\tau) &= F_3G_3(\tau) + E_5G_5(\tau), \\
			g_{n\geq 4}(\tau) &= F_nG_n(\tau) + E_{n+2}G_{n+2}(\tau) + E_nG_{n-2}(\tau),
		\end{align}	
	\end{subequations}
	where $\{E_n,F_n\}$ are constants to be determined from the boundary conditions.
	
	\subsubsection{Boundary conditions for $u_\tau$}
	The impenetrability condition implies 
	\begin{equation}
		\frac{\partial \psi}{\partial \zeta}(\tau_0,\zeta) = 0,
	\end{equation}
	i.e.~$\psi$ is constant on the boundary. Since  $\psi$ is zero at the poles ($\psi(\tau_0,\zeta) = 0$) from the property $G_{n\geq2}(\zeta = \pm 1) = 0$ of the Gegenbauer functions, $\psi$ is zero everywhere on the boundary, yielding our first boundary condition for the $g_n$'s,
	\begin{equation}\label{BC1}
		g_n(\tau_0) = 0.
	\end{equation}
	
	\subsubsection{Boundary conditions for $u_\zeta$}
	
	We write the tangential velocity boundary condition in terms of $\psi$ and use the orthogonality relations 
	\begin{equation} 
		\int_{-1}^1\frac{G_nG_m}{1-\zeta^2}\,\mathrm d\zeta = \frac{2}{n(n-1)(2n-1)}\delta_{mn}, \qquad n,m > 2,
	\end{equation}
	and the relation $(1 - \zeta^2)^{1/2}P_l^1(\zeta) = -l(l + 1)G_{l+1}(\zeta)$ between the Gegenbauer functions and the associated Legendre polynomials $P_l^1$  to obtain
	\begin{equation}
		\frac{\mathrm d g_n}{\mathrm d \tau}(\tau_0) = c^2\int_{-1}^{1}\sqrt{\tau_0^2-\zeta^2}P_{n-1}^1(\zeta)v_s(\zeta)\,\mathrm d\zeta.
	\end{equation}
	This motivates the expansion of the prescribed boundary flow $v_s$ as
	\begin{equation}\label{vsexpansion}
		v_s(\zeta) =  \frac{\tau_0}{\sqrt{\tau_0^2 - \zeta^2}}\sum_{n \geq 1}B_n P_n^1(\zeta).
	\end{equation} 
	The tangential velocity boundary condition may then be expressed as
	\begin{equation}\label{BC2}
		\frac{\mathrm d g_n}{\mathrm d \tau}(\tau_0) = \tau_0c^2n(n-1)B_{n-1}.
	\end{equation}	
	The coefficients $B_n$ are determined by the prescribed boundary flow $v_s$, and explicit expressions follow from Eq.~\eqref{vsexpansion}:
	\begin{equation}\label{eq:bn}
		B_n = \frac{n + \frac{1}{2}}{n(n+1)\tau_0}\int_{-1}^{1}v_s(\zeta)\sqrt{\tau_0^2 - \zeta^2}P_n^1(\zeta)\,\mathrm d\zeta.
	\end{equation}
	Note that on the boundary $\tau = \tau_0$ of the spheroid, $\zeta$ is simply a rescaled axial position, i.e.~$\zeta = z/b_z$.

	\subsubsection{Solving the boundary conditions for $\{E_n,F_n\}$}\label{section:solve}
	
	As is standard in many Stokes flow problems~\cite{happel1965}, these boundary conditions yield an infinite system of linear equations for an infinite number of unknowns $\{E_n,F_n\}$, which we may solve order by order and thereby, from Eqs.~\eqref{u},\eqref{psi}, and \eqref{g}, determine the full flow field inside the spheroid. 
	
	We solve the two linear equations from the boundary conditions, 
	Eqs.~\eqref{BC1} and \eqref{BC2}, for $n = 2$ together with   Eq.~\eqref{g2} for $g_2$ for the two unknowns $F_2$ and $E_4$. At $n = 4$, we similarly have two equations involving $\{E_4, E_6, F_4\}$, which we may again solve since $E_4$ is known from the previous set of equations; this procedure may be continued to determine $\{E_n,F_n\}$ for all even $n$.  An analogous method determines $\{E_n,F_n\}$ for odd $n$, and we may thus calculate the coefficients to arbitrary order. 
	
	\subsection{Flow fields}\label{section:modes}
	
	In the rest of this paper,  we truncate our solution at $n = 14$, since higher orders give minimal gains in accuracy; making  instead, for instance, the choice  $n = 25$ leads to a difference of 0.1\% or less (except at points where the flow is already almost zero).

	\begin{figure}[t!]
		\includegraphics[width=0.9\textwidth]{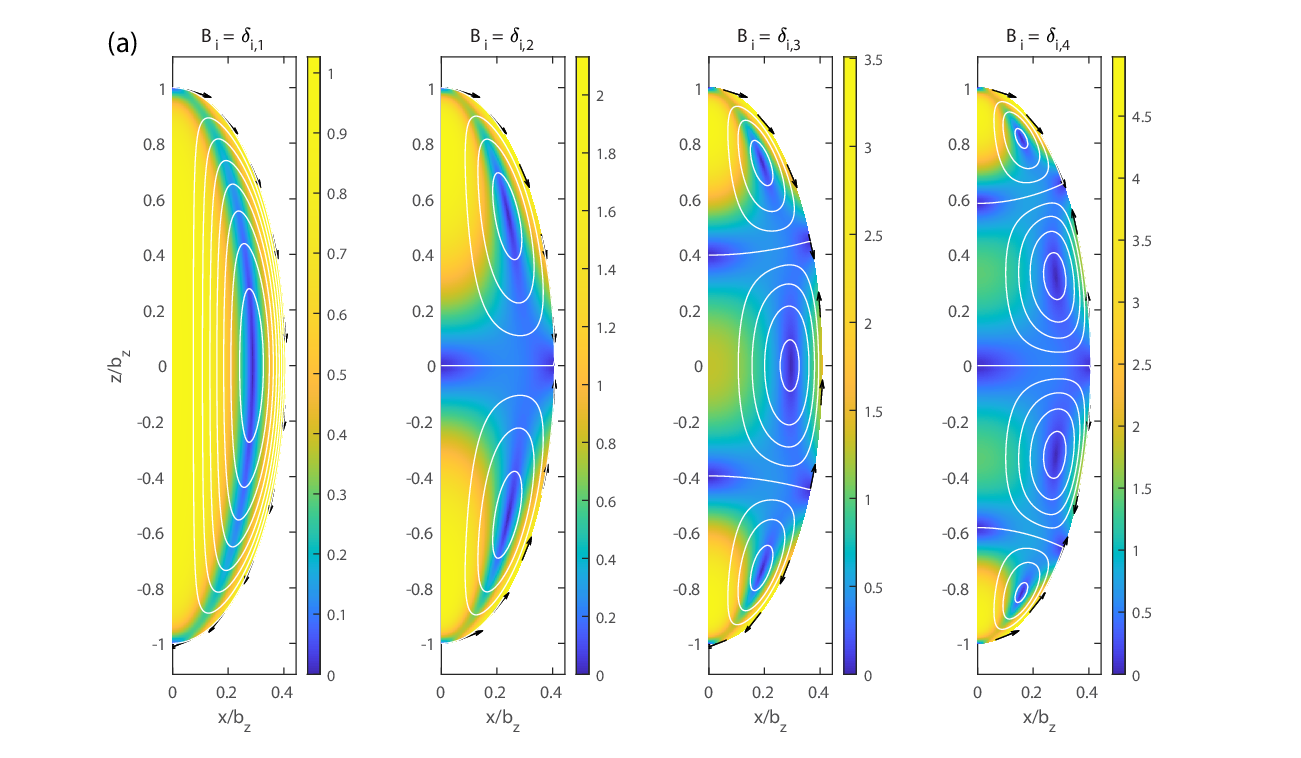}
		\includegraphics[width=0.9\textwidth]{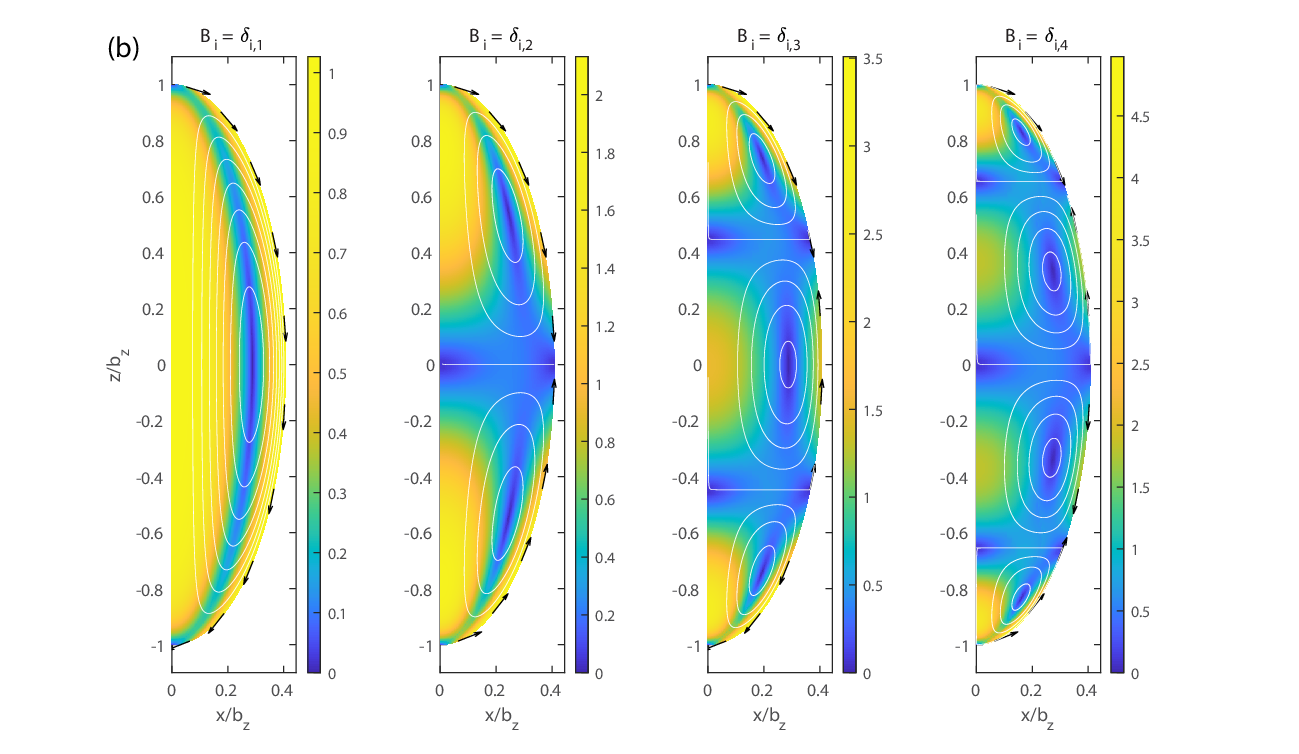}
		\caption{
			(a) Flow fields corresponding to the first four modes in the series expansion in Eq.~\eqref{vsexpansion} of the tangential slip velocities using the exact solution from \S~\ref{section:maths}. Colour indicates the magnitude $|\mathbf u(\mathbf x)|$ of the velocity field. White lines are streamlines. Black arrows at the boundary indicate tangential velocities. 
			(b) Same result as derived by the long-wavelength model of \S~\ref{section:lubrication}.}\label{fig:modes}
	\end{figure}

	We illustrate in Fig.~\ref{fig:modes} the first few modes in the geometry relevant to  \textit{Drosophila} embryos, with  $b_x = 110$~\textmu{m} and $b_z = 270$~\textmu{m}, corresponding to $\tau_0 = 1.095$~\cite{deneke2019}. We show in Fig.~\ref{fig:modes} the 
	flow fields corresponding to the first four modes,  i.e.~the cases with $B_{i} = \delta_{im}$ for $m = 1,2,3,4$.  The flow field corresponding to the $m^{\text{th}}$ mode is seen to display $m$ vortices inside the cell.

	%%%%%%%%%%%
	\section{Optimal nuclear transport by cortex-driven cytoplasmic flows in \textit{Drosophila} embryo}\label{section:application}

	We now use the full flow solution we have computed to  study the specific features of flows and transport in the \textit{Drosophila} embryo. After fertilisation, the nucleus of the \textit{Drosophila} embryo undergoes 13 rounds of cell division, referred to as ``cell cycles''~\cite{farell2014, rabinowitz1941}.  
	During the first three cell cycles, the nuclei produced by cell division undergo  very little movement. However, during cell cycles 4 to 6, the nuclei   are seen experimentally to spread along the long axis of the embryo~\cite{zalokar1976}.  This so-called ``nuclear spreading'' is driven by cortical contractions and the resulting cytoplasmic flows~\cite{deneke2019}. Once the nuclei have achieved a uniform distribution along the AP axis, from cell cycle 7 onwards, cytoplasmic flows and nuclear movements become significantly smaller. In this section, we  use our mathematical solution to investigate the flow-based nuclear spreading in cell cycles 4-6. Specifically, we show theoretically how our  mathematical model can reveal  near-optimal cortical flows for nuclear spreading.

	\subsection{Cytoplasmic flows during cell cycles 4 to 6}
	
\begin{figure}[t!]
		\includegraphics[width=\textwidth]{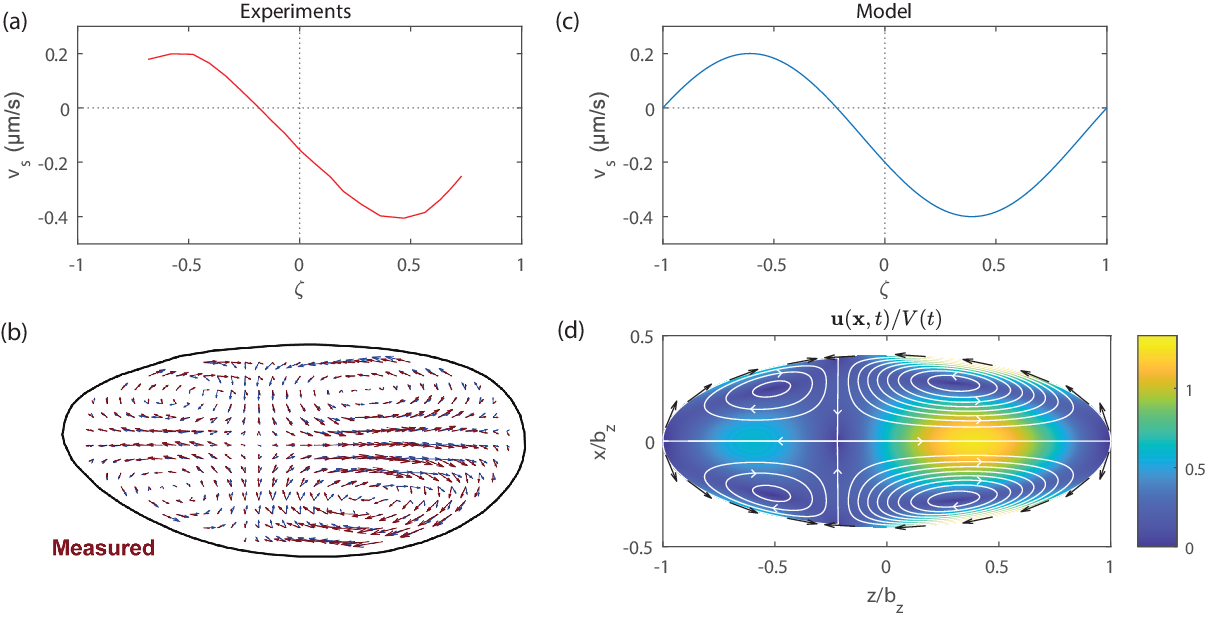}
		\caption{Modelling cytoplasmic flows during cell cycles 4 to 6 of \textit{Drosophila} development. (a,b) Experimental measurements of (a) the cortical flow profile at the contraction peak of cell cycle 6, redrawn from Ref.~\cite{deneke2019} and (b) cytoplasmic flows inside the \textit{Drosophila} embryo (red arrows), reprinted from \textit{Cell},  \textbf{177}, Deneke \textit{et al.}, ``Self-Organized Nuclear Positioning Synchronizes the Cell Cycle in \textit{Drosophila} Embryos", 925, Copyright (2019), with permission from Elsevier \cite{deneke2019}. (c) The cortical flow $v_s$ as prescribed in the mathematical model, Eq.~\eqref{eq:modelflow}. (d) The cytoplasmic flows obtained from the analytical solution. Black arrows indicate cortical flows at the boundary, streamlines are illustrated in white, and the colour map indicates flow speed. }\label{fig:flow}
	\end{figure}

	The \textit{Drosophila} embryo is  approximately a prolate spheroid with semi-minor axis  $b_x = 110$~\textmu{m} and semi-major axis $b_z = 270$~\textmu{m} (see Figs.~\ref{fig:intro} and~\ref{fig:flow}b)~\cite{deneke2019}.  Mapping {onto our modified prolate spheroidal coordinates}, these dimensions correspond to $c = \sqrt{b_z^2 - b_x^2} = 246.6$~\textmu{m} and $\tau_0 = b_z/c = 1.095$.

	To model the cortical flow in space and time, we use the experimentally measured cortical flows in cell cycle 6 of the \textit{Drosophila} embryo~\cite{deneke2019}, reproduced in Fig.~\ref{fig:flow}a, with full flow field shown in Fig.~\ref{fig:flow}b. We model this flow distribution with a shifted sine profile; this approximates the experimental profile well {in the range of $\zeta$ in which measurements are taken, and extrapolates to zero at the poles, which we would expect since there are no sources/sinks of cortical matter at the poles.}  {Real-life cortical flows have a complex space- and time-dependence, but to a good approximation may be modelled with a time-dependent amplitude $V(t) \geq 0$ and a spatial profile independent of time, given by the fitted shifted sine function.} The boundary conditions for the flows are therefore given by
	\begin{equation}\label{eq:modelflow}
		v_s(\zeta, t) = V(t)\left[-\sin\left(\pi\zeta + \arcsin\frac{1}{3}\right) - \frac{1}{3}\right].
	\end{equation}
	This boundary profile, with $V(t)$ taken to be 0.3~\textmu{m}/s at the cortical contraction peak of cell cycle 6, is illustrated in Fig.~\ref{fig:flow}c, closely matching the experimental profile from  Fig.~\ref{fig:flow}a. 
	
	Using Eq.~\eqref{eq:modelflow} as boundary condition, the flow inside the spheroid is computed as outlined in \S\ref{section:maths}. First the $B_n$'s are evaluated using Eq.~\eqref{eq:bn}. Next the boundary condition in Eqs.~\eqref{BC1} and \eqref{BC2}, with the $g_n$'s as defined in Eq.~\eqref{g}, are solved for the coefficients $\{E_n,F_n\}$ as   explained in section \S\ref{section:solve}. This determines the streamfunction (Eq.~\ref{psi}) and in turn the flow field (Eq.~\ref{u}).
	
	The  experimentally measured flow field is shown as a vector field in Fig.~\ref{fig:flow}b and we plot our theoretical prediction in Fig.~\ref{fig:flow}d. The   analytical model matches experiments well, capturing the cortical flows directed towards the middle of the embryo  that then drive cytoplasmic flows along the long axis towards the poles. The model also   reproduces the fore-aft asymmetry of the flow (resulting from the asymmetric boundary flows evident in  Figs.~\ref{fig:flow}a and \ref{fig:flow}c) and   the four-vortex structure of the bulk flow.

	\subsection{Nuclear transport}
	
	\subsubsection{Modelling}
	
	One purpose of the cytoplasmic flows shown in Fig.~\ref{fig:flow} in cell cycles 4 to 6 is to distribute the cell nuclei along the AP axis of the embryo.  These nuclei are subject to both advective transport from the cytoplasmic flows and to Brownian motion.
	
	First, let us show that  Brownian motion can be neglected. Each nucleus produced by  cell division has a diameter of approximately 7~\textmu{m}. Assuming a cytoplasm at room temperature with an effective  viscosity equal to that of water, a simple application of the Stokes-Einstein relationship allows us to estimate the diffusion constant of each nucleus as approximately 0.06 \textmu$\text{m}^2~\mathrm s^{-1}$. Over the course of cell cycles 4 to 6 (roughly 30 minutes), the contribution from diffusion to the root mean squared  displacement of a nucleus  is thus estimated to be at most on the order of 10~\textmu{m}, much smaller than the typical size of the embryo.  Further, the true diffusive displacement is likely to be significantly smaller because the cytoplasm is known to be more viscous than water~\cite{swaminathan1997}  and because confinement effects within the  \textit{Drosophila} embryo  decrease the Stokes mobility of the nuclei~\cite{kimbook}.  It is thus appropriate to assume that  the nuclei do not diffuse appreciably over the time scale of cell cycles and we may neglect Brownian diffusion in their transport during cycles 4 to 6.

	In the beginning of cell cycle 4, a cloud of nuclei of radius approximately $60$~\textmu{m} is found within the anterior half of the embryo, and is subsequently spread along the long axis~\cite{sullivan1995, deneke2019}. In order to model transport, we  simulate the advection of a cloud of $N$ passive tracer particles 
	drawn from a uniform distribution over a sphere of radius 60~\textmu{m} centered at the stagnation point of the cytoplasmic flows, i.e.~at the point $\zeta = -2\arcsin(1/3)/\pi \approx -0.2163$ on the long axis (see Fig.~\ref{fig:flow}a).  The  particles are  then simply   advected by the  cytoplasmic flow, $\mathbf u$. Specifically, the rate of change of the position $\mathbf x_i$ of the $i^{\text{th}}$ particle is  equal to the instantaneous flow 
	$\mathbf u(\mathbf x_i,t)$ at position $\mathbf x_i$ and time $t$, i.e.~it satisfies the differential equation
	\begin{equation}\label{eq:adv}
		\frac{\mathrm d\mathbf x_i}{\mathrm d t} = \mathbf u(\mathbf x_i,t).
	\end{equation}

	\subsubsection{Integral measure of cortical contractions}
	Under these modelling assumptions,  nuclear transport inside the embryo can be shown to depend 	 on a single integral measure of the cortical contractions. Indeed,  Stokes flows (i.e.~flows of Newtonian fluids in the absence of inertia) have no dependence on time other than through the boundary conditions~\cite{kimbook}.  Since the advection equation for the nuclei, Eq.~\eqref{eq:adv}, is deterministic, the state of the system is fully determined by the  quantity $\chi$ defined as the integral in time of the amplitude of cortical contractions $V(t)$ from Eq.~\eqref{eq:modelflow}, i.e.
	\begin{equation}
		\chi := \int_0^{t}V(t')\,\mathrm dt'.
	\end{equation}
	This quantity  $\chi$ has units of distance, and can be interpreted as (approximately) the total distance travelled by material points along the embryo boundary; it can therefore be  thought of as a rescaled time, where $\chi$ increases with time {(provided that $V(t)$ is positive)}, with the  precise  details of $V(t)$ no longer important. {In the following simulations, we therefore use $\chi$ rather than $t$ as the time variable.} 
	
\begin{figure}[t!]
		\includegraphics[width=0.85\textwidth]{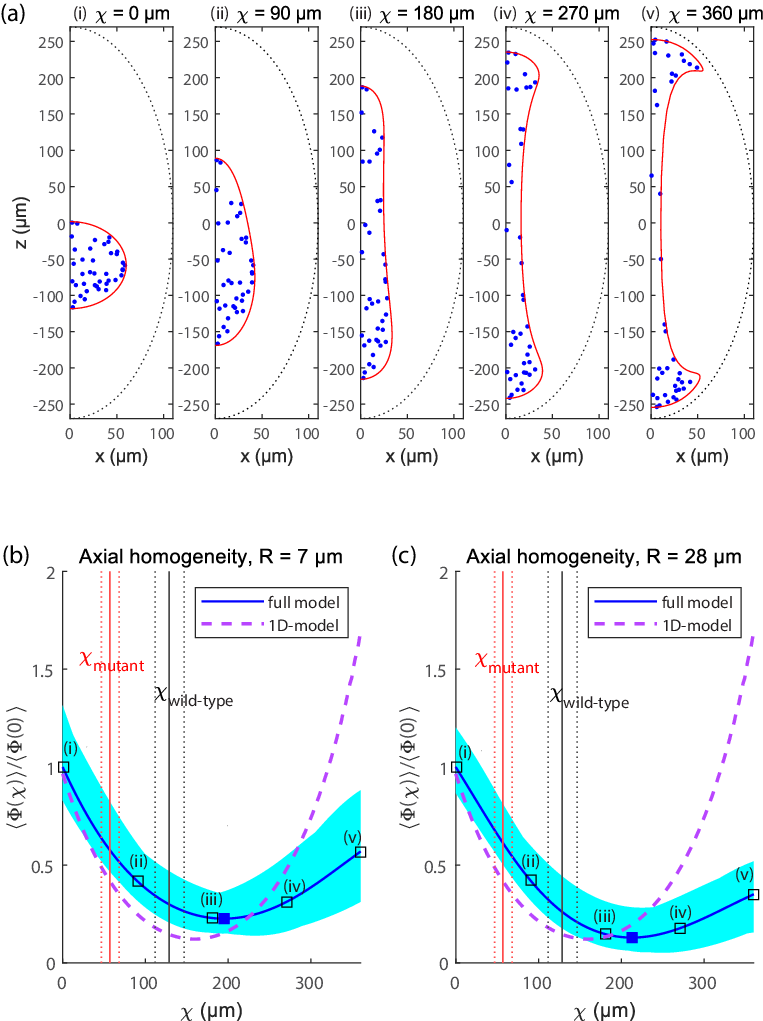}
		\captionof{figure}{Caption on  next page}
		\label{fig:PHI}
	\end{figure}

	The simulation results are shown in Fig.~\ref{fig:PHI}a in the case $N=40$, and illustrate how the  nuclei, initially arranged within a  60~\textmu{m} sphere,  are progressively spread by cytoplasmic flows along the axis of the embryo as the value of  $\chi$  increases.  We further plot in red how the boundary of the initial  sphere is distorted by the flows. Strikingly, we see from these computational   results  that the bulk flows (driven by cortical contractions) initially facilitate the axial spreading of the nuclei; however, beyond some critical value of $\chi$,  applying flows further appears to be counter-productive, as they produce a `neck' scarce in particles and accumulate a disproportionate number of nuclei near the poles of the embryo, eventually resulting in recirculation of nuclei towards the centre along the boundary.

	\quad	\clearpage
	{FIG. 5: Optimal spreading of nuclei by cortical contractions. 
		(a) Advective transport of a cloud of tracer particles (blue) as models for nuclei  initially at random positions in a sphere of radius 60~\textmu{m}, at different values of the rescaled time $\chi$, using 	the exact flow solution from \S~\ref{section:maths} . Advection of the sphere's boundary is illustrated in red. (b,c) Blue solid line: Measure of axial homogeneity against $\chi$, obtained by computing the
		mean variance, $\langle \Phi(\chi)\rangle$, averaged over 100 simulations of nuclear transport, relative to its mean initial value   $\langle\Phi(0)\rangle$, {using the density kernel length scales $R = 7$~\textmu{m} (b) and  $R = 28$~\textmu{m} (c)}; shaded area indicates the minimum and maximum value of $\Phi$ at each $\chi$. Black open squares labelled (i)-(v) correspond to the values of $\chi$ in the snapshots in (a). Blue diamond indicates the optimal value of $\chi$ (which minimises  $\langle \Phi(\chi)\rangle$). {Black and red  lines indicate estimates of the average (solid lines)  value of $\chi$ realised in wild-type and mutant (PPI-het) \textit{Drosophila} embryos respectively and at the 95\% confidence  levels (dotted lines).} The purple dashed line indicates axial homogeneity as predicted by a reduced one-dimensional   model of transport (\S~\ref{section:lubrication}), with the  purple diamond indicating the optimal $\chi$ as predicted by that reduced model.}
	
	\bigskip
	
	\hrule
	
	\bigskip

	\subsubsection{Optimal cytoplasmic transport and cortical contractions}
	\label{sec:optimal}
	
	These numerical results therefore seem to indicate that an optimum $\chi$ exists to ensure a uniform axial spreading of the nuclei. To make this observation more quantitative, we  introduce a mathematical measure of axial homogeneity in the distribution of nuclei. Specifically, we define a continuous density function $\rho(z,\chi)$ of nuclei along the long $z$ axis of the embryo by placing a Gaussian kernel at the centre of each nucleus. With $N$ nuclei, we define 
	\begin{equation}\label{def:density}
		\rho(z, \chi) := \frac{2}{NR\sqrt{2\pi}}\sum_{i = 1}^N e^{-[z - z_i(\chi)]^2/2R^2},
	\end{equation}
	where $z_i(\chi)$ is the position of the $i^{\text{th}}$ nucleus along the long axis at rescaled time $\chi$ and where {the length scale $R$ controls the width of the Gaussian kernel}. The homogeneity of particle distribution along the axis may then be measured by the variance  $\Phi$ of the density, defined as
	\begin{equation}
		\Phi(\chi) := \int_{-b_z}^{b_z}[\rho(z,\chi) - \bar\rho(\chi)]^2\,\mathrm dz
	\end{equation}
	where $\bar\rho(\chi) = \frac{1}{2b_z}\int_{-b_z}^{b_z}\rho(z,\chi)\,\mathrm dz$ is the average density; note that the variance is only a function of $\chi$, i.e.~the rescaled time.

	With these definitions, we run 100 simulations (i.e.~100 sets of random  initial conditions for the model nuclei) and plot the mean value $\langle \Phi(\chi)\rangle$, relative to the average initial value in the simulations $\langle\Phi(0)\rangle$, {for two biologically relevant values for the Gaussian kernel length scale: (i) $R = 7$~\textmu{m}, which is the approximate diameter of a nucleus (results shown in Fig.~\ref{fig:PHI}b, solid line); and (ii) $R = 28$~\textmu{m}, the mean nuclear separation distance determined by microtobule aster migration (an aster is a radial array of microtubules attached to a centrosome)  during cell division~\cite{telley2012} (results shown in Fig.~\ref{fig:PHI}c). In each plot,} the shaded regions indicate the minimum and maximum over the 100 simulations of $\Phi(\chi)$ at each $\chi$. These quantitative results confirm the intuitive observation gleaned from the images in Fig.~\ref{fig:PHI}a: the optimal axial homogenisation of the nuclei happens for a well-defined value of the integral measure of cortical contractions, $\chi$.  Specifically, this  optimal spreading is predicted by our numerical simulations to take place at {$\chi \approx 195$~\textmu{m} using $R = 7$~\textmu{m} and $\chi \approx 213$ \textmu{m} using $R = 28$~\textmu{m}.} (Here, and in what follows, we give all figures for $\chi$ to the nearest integer value.)
	
	\begin{figure}[t!]
		\includegraphics[width=\textwidth]{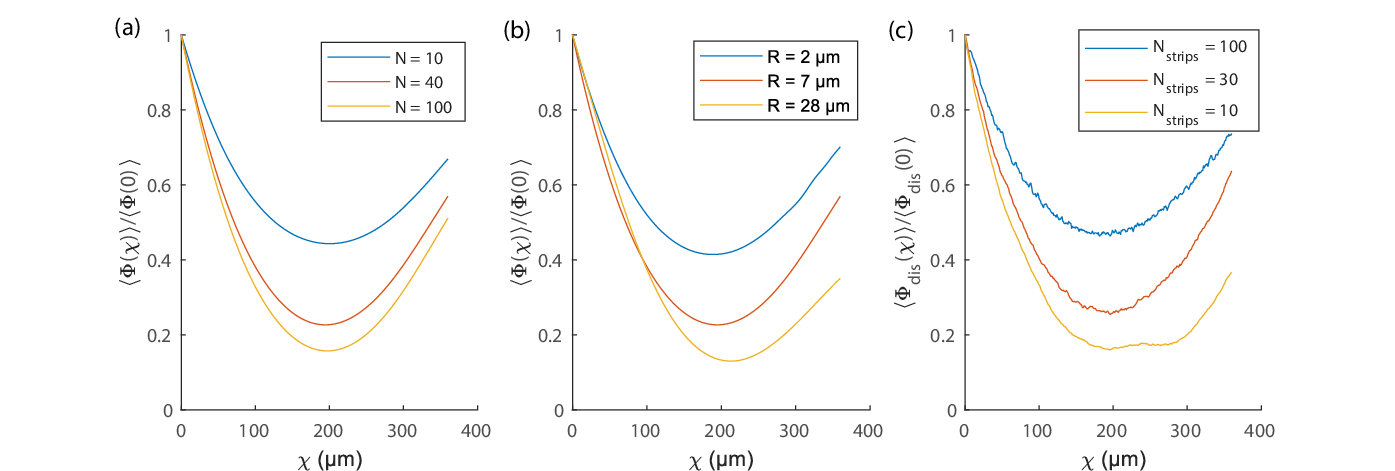}
		\caption{Dependence of optimal cortical forcing on modelling parameters. 
			(a) Average axial homogeneity, $\langle\Phi(\chi)\rangle$, with {$R = 7$~\textmu{m}}, plotted for three different numbers $N$ of particles advected by the flow. 
			(b) Average axial homogeneity, $\langle\Phi(\chi)\rangle$, for $N = 40$ particles, plotted  for three different values of the Gaussian kernel length scale, $R$. 
			(c) Alternative approach where a different, discrete, measure of axial homogeneity is introduced, $\langle\Phi_{\text{dis}}(\chi)\rangle$ (see text), with results shown for $N = 40$ particles and three values of the number  $N_{\text{strips}}$ of strips. 		}\label{fig:supp}
	\end{figure}

	Importantly, the  presence of this optimal spreading is very robust to changes in  our modelling parameters. Varying the number of particles $N$ advected by the flow produces essentially no  difference, as illustrated in Fig.~\ref{fig:supp}a where the result for 	$R = 7$~\textmu{m} and $N=40$ ({optimum at $\chi \approx 195$~\textmu{m}}) is compared with those for $N=10$ and 100 {(optima at $\chi \approx 200$~\textmu{m} and 197~\textmu{m}, respectively)}. {Changing the  length scale $R$ associated with the Gaussian kernel in the definition of the continuous density,   Eq.~\eqref{def:density}, also has no qualitative effect on our results, as illustrated in      Fig.~\ref{fig:supp}b, and  the optimal   value of $\chi$ is seen to only  increase weakly with $R$: our model  predicts optima at {$\chi \approx 189$~\textmu{m}, 195~\textmu{m} and 213~\textmu{m} for the choices $R = 2$~\textmu{m}, 7~\textmu{m} and 28~\textmu{m},  respectively.} 
		
		Note that alternatively, we could   use a `discrete' approach to 
		quantify axial homogeneity. To do this, we divide the embryo into $N_{\text{strips}}$ strips of equal width in the direction perpendicular to the long axis and define 
		$\Phi_{\text{dis}}(\chi)$ as the variance over $i$ of the number $n_i$ of particles located in strip $i$ at the  rescaled time $\chi$.   This  average variance,  $\langle\Phi_{\text{dis}}(\chi)\rangle$, over 100 simulations, is plotted in Fig.~\ref{fig:supp}c as a function of $\chi$ for the choices $N_{\text{strips}}=100$, 30 and 10 (in all cases, we solve for the transport of $N=40$ particles). Here also we see a clear minimum of the variance associated with optimal spreading, at $\chi \approx 196$~\textmu{m}, 196~\textmu{m} and 195~\textmu{m}, respectively.

		\subsubsection{Comparison with experiments}\label{sec:comp}
		
		How do these theoretical predictions compare with the  value of $\chi$ realised in real \textit{Drosophila} embryos?  This may be estimated from  the experimental results in 
		Ref.~\cite{deneke2019}.

		We first estimate the cortical flow amplitude $V(t)$  for a wild-type embryo from an experimentally measured time series of the RMS cytoplasmic flow field (reproduced in Fig.~\ref{fig:vmodel}a, dark blue line) and the cortical velocity profile in the contraction phase of cell cycle 6 (reproduced in Fig.~\ref{fig:flow}a).

		We may assume that the transport is dominated by the three large cortical contraction peaks in the three cell cycles (see peaks in Fig.~\ref{fig:vmodel}a, dark blue line), and can neglect contributions from the intervals between the peaks, {since intervals with no flow lead to unchanged values of $\chi$}. Because our model assumes Stokes flows in the Newtonian limit, the bulk cytoplasmic flows depend linearly on the cortical flows and we may therefore rescale RMS cytoplasmic velocities (Fig.~\ref{fig:vmodel}a) into  cortical flow amplitudes. Therefore we sample approximately 20 points from each contraction peak in the time series of the RMS cytoplasmic speeds (Fig.~\ref{fig:vmodel}a, dark blue line) and rescale this data into cortical flow amplitudes, with the scaling factor chosen to obtain a cortical flow amplitude of  0.3~\textmu{m}/s at the contraction peak of cell cycle 6; recall that the value $V(t) = 0.3$~\textmu{m}/s in our model for the cortical flow, Eq.~\eqref{eq:modelflow}, is the one that gives close agreement with experiments (Fig.~\ref{fig:flow}c). 
		We then fit a sum of three Gaussians to the rescaled data to obtain a mathematical estimate as $V(t) = \Sigma_{i = 1}^3a_i\exp[-(t - b_i)^2/2c_i^2)]$ for the time-variation  of the cortical flow amplitude; the resulting model for  the function $V(t)$  is illustrated in Fig.~\ref{fig:vmodel}b.

		\begin{figure}[t!]
			\includegraphics[width=0.7\textwidth]{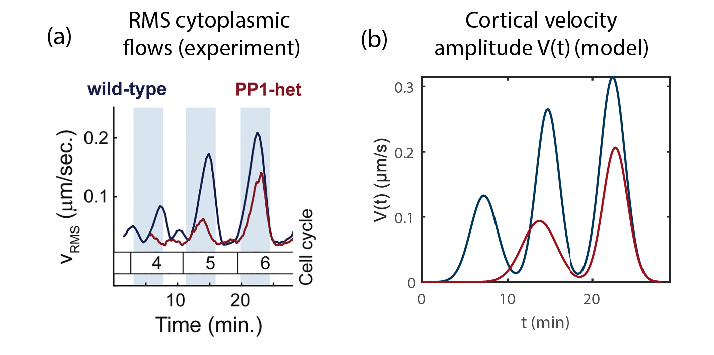}
			\caption{Estimation of $\chi$ from experiments. 
				(a) RMS cytoplasmic flows measured experimentally for a wild-type (blue) and mutated (red) \textit{Drosophila} embryo; reprinted from \textit{Cell},  \textbf{177}, Deneke \textit{et al.}, ``Self-Organized Nuclear Positioning Synchronizes the Cell Cycle in \textit{Drosophila} Embryos", 925, Copyright (2019), with permission from Elsevier \cite{deneke2019}. (b) Model for cortical flow amplitude $V(t)$ obtained by fitting a sum of Gaussians to the large contraction peaks in cell cycles in (a) and rescaling appropriately to convert RMS cytoplasmic speeds to cortical flow amplitudes.
			}\label{fig:vmodel}
		\end{figure}
	
		This quantitative model for the cortical flow amplitude $V(t) $ can then be integrated in time, leading to an  estimate for the   value of $\chi$ occuring in \textit{Drosophila} embryos.  Using a fit of three Gaussians at the 95\% confidence interval level, we obtain the  estimated range for $\chi$ of $ 112$~\textmu{m}--$147$~\textmu{m}, with mean value $\langle \chi \rangle \approx 129$~\textmu{m}. {These mean values  are illustrated in  Figs.~\ref{fig:PHI}b and \ref{fig:PHI}c with a black vertical line, and dotted lines are used as error bars for the range of $\chi$ at the 95\% confidence level.}  
		Remarkably, we see from these results   that the magnitude of the cortical contractions in experiments is close to the optimal value predicted theoretically by our simple mathematical  model;  real-life cortical flows appear thus to be  near-optimal to ensure axial spreading of nuclei.

		In the experiments of Ref.~\cite{deneke2019}, an embryo with mutations (named `PP1-heterozygous' or `PP1-het')  had dampened cytoplasmic flows; this leads to non-uniform nuclear spreading and has been shown to cause improper control and asynchrony in the following cell cycles, impairing the development process. The value of $\chi$  realised for this mutant  can be similarly estimated by fitting a sum of two Gaussians  $V(t) = \Sigma_{i = 1}^2A_i\exp[-(t - B_i)^2/2C_i^2)]$ (Fig.~\ref{fig:vmodel}) to the time series of the  cortical velocity amplitude rescaled from the experimentally measured RMS cytoplasmic velocity in this embryo (Fig.~\ref{fig:vmodel}a). For this embryo, the value of $\chi$ realised is $\chi \approx 57$~\textmu{m}, with ranges of error estimated to be 47~\textmu{m}--68~\textmu{m} at {the 95\% confidence level of the fit}. These values are illustrated  {in  Figs.~\ref{fig:PHI}b and \ref{fig:PHI}c as red solid line (mean) and dotted lines (95\% error bars)}. We see that dampening the flows leads to a corresponding decrease in the extent of nuclear spreading achieved, captured quantitatively by a significantly smaller value of $\chi$.

		{
			\subsubsection{Impact of cell division}			 
			Our model of transport with passive tracers neglects the fact that the nuclei divide in each cell cycle. However, incorporating cell division and the resulting microtubule aster  migration-induced nuclear separation would lead to additional spreading and homogenisation, allowing the system to achieve optimal spreading at a lower value of $\chi$, and thus bringing our theoretical optimum even closer to the experimental estimates.}

		{To show this more quantitatively, we focus on the results   in  Fig.~\ref{fig:PHI}d using the kernel length scale $R = 28$~\textmu{m}, since  this is the  nuclear separation length reported experimentally \cite{telley2012}  and is thus   the appropriate length scale to use when dividing nuclei are considered. Indeed, in order for $\Phi$ to be a measure of homogeneity that can be interpreted meaningfully across different numbers of nuclei in the embryo at different cell cycles, the single-peak density function associated with a single isolated nucleus should remain approximately the same after it divides into two (in the absence of flow). Using $R = 28$~\textmu{m} ensures this, while using a significantly smaller values of $R$ results in a qualitatively different two-peak density function for the daughter nuclei.}

		{We  performed simulations of cell division in the absence of flow, modelling the nuclear separation as a repulsive force field  of range $28$~\textmu{m} around each nucleus. Specifically, 32 nuclei distributed uniformly in a spheroidal cloud of semi-major axis 140 \textmu{m} and semi-minor axis 40 \textmu{m}, a configuration which approximates the beginning of cell cycle 6, are allowed to divide into 64 nuclei. The values of $\langle \Phi \rangle$ before and after cell division are computed over 100 simulations, and we find a decrease of approximately a 6\% in the value $\Phi$ due to one round of cell division.} 
		
		{We expect however that the decrease in $\langle\Phi\rangle$ due to the three cell divisions occurring in cell cycles 4-6 will be less than 18\% (i.e.~three times 6\%) of the initial value at $\chi = 0$, primarily because the later cell divisions occur at values of $\Phi$ which have already been lowered by cytoplasmic flow. Furthermore, the estimate of $6$\% was obtained from a configuration which emulates the beginning of cell cycle 6, when the nuclei are most numerous and closely packed and when therefore the spreading induced by physical repulsion would be a maximum. We  estimate that incorporating cell division into the full model in Fig.~\ref{fig:PHI}c would contribute a decrease in $\langle \Phi \rangle$ relative to $\langle \Phi(\chi = 0)\rangle$ of around 10\%. The optimum value of $\chi$ with cell division included can therefore be estimated as the value of $\chi$ at which $\langle \Phi (\chi)\rangle/\langle \Phi (0)\rangle$ attains a value 0.1 higher than its minimum value (without cell division). This value is $\chi \approx 140$ \textmu{m}, which is significantly closer to the experimental estimate.
		}

		\section{Long-wavelength theory}
		\label{section:lubrication}

		Thus far, we have  investigated flow-based transport in the \textit{Drosophila} embryo using our analytical solution, written as an infinite series of spheroidal harmonics, for the cytoplasmic flow. This flow solution is exact, but    is restricted to cell shapes that are exactly spheroidal.  In this section we use a long-wavelength assumption (also known as lubrication theory in the context of fluid mechanics) to derive a much simpler solution for the flow, which nonetheless approximate the exact flows remarkably well. Importantly, this approach is applicable to any elongated (axisymmetric) shape, and is thus not restricted to spheroidal cells.
		
		We  work in cylindrical coordinates $(r,\phi,z)$, with $z$ aligned with the long axis of the cell and denote the radius of the cell by $R(z)$. 		The key assumption is that length scales in the longitudinal direction are much larger than radial length scales~\cite{leal2007}; this means the  long-wavelength model is valid in the limit $|R'(z)|\ll 1$. We seek a solution for the cytoplasmic flow $\mathbf u(r,z,t) = u_z(r,z,t)\mathbf e_z + u_r(r,z,t)\mathbf e_r$. Under the long-wavelength assumption, the cell geometry is locally cylindrical, and there are no radial pressure gradients to leading order. The leading-order longitudinal flow is therefore pressure-driven pipe flow subject to a slip velocity $u_z = U(z,t)$ at $r = R(z)$,  where 
		\begin{equation}
			U(z,t) = {\bf e}_z\cdot {\bf v}_s\big|_{ \zeta = {z}/{b_z}} 
		\end{equation}
		is the component along the $z$ direction of the cortical velocity $\mathbf v_s$. Since the ends of the cell are closed and the fluid is incompressible, there is no net mass flux through any cross-section, i.e.~$\int_0^R u_z r\,\mathrm dr = 0$.

		{	We now derive
			the slowly-varying  flow solution which satisfies these conditions. The radial component of the Stokes equations for an incompressible fluid of shear viscosity $\mu$ in the lubrication limit reads $\partial p/\partial r = 0$, where $p$ is the dynamic pressure. Note that $\partial p/\partial \theta = 0$ follows from the assumption of axisymmetry.  This allows us to straightforwardly integrate  the axial component  of the Stokes equations in cylindrical coordinates, $\mu \nabla^2 u_z = {\partial p}/{\partial z}$,  to yield the general solution
			\begin{equation}\label{eq:u}
				u_z = \frac{1}{4\mu}\left(\frac{\partial p}{\partial z}\right) r^2 + A \ln r + B,
			\end{equation}
			where $A$ and $B$ are integration constants. Regularity of the flow at the origin requires $A = 0$. The unknown pressure gradient $\partial p/\partial z$ and the constant $B$ may then be determined by imposing  (i) the slip velocity boundary condition $u_z = U(z,t)$ at $r = R(z)$ and (ii) zero total mass flux throughout the cell, $\int_0^R u_z r\,\mathrm dr = 0$, yielding the final long-wavelength solution 
			\begin{equation}\label{eq:u}
				u_z(r,z,t) = {U(z,t)}\left(2\frac{r^2}{R^2} - 1\right),
			\end{equation}
			for the longitudinal (axial) velocity in cortically driven cytoplasmic flow.} 
		Integrating the incompressibility condition, $\nabla \cdot \mathbf u = 0$, for the radial velocity $u_r$ then yields 
		\begin{equation}\label{eq:v}
			u_r(r,z,t) = \frac{1}{2}\frac{\partial U(z,t)}{\partial  z} r\left[1 - \left( \frac{r}{R(z)}\right)^2 \right] +U(z,t)\frac{\mathrm dR(z)}{\mathrm dz}\left( \frac{r}{R(z)}\right)^3.
		\end{equation}
		Note that regularity at $r = 0$ required the integration constant to vanish, and the solution then satisfies the no-penetration boundary condition on the cell wall,  
		$(u_z\mathbf e_z + u_r\mathbf e_r)\cdot\mathbf n = 0$,
		where $\mathbf n~\propto~(\mathrm dR/\mathrm dz~\mathbf e_z~-~\mathbf e_r)$ is a vector normal to the boundary. Together with the slip velocity $u_z = U(z,t)$, this is equivalent to the original no-slip  boundary condition in Eq.~\eqref{eq:no-slip},  $\mathbf u = \mathbf v_s$.		The long-wavelength solution in Eqs.~\eqref{eq:u} and \eqref{eq:v} satisfies therefore the boundary conditions of the exact problem.

		To illustrate the prediction of this model, we reproduce in  Fig.~\ref{fig:modes}b   the first four  spheroidal modes from the exact solution in \S~\ref{section:maths} but computed  using the long-wavelength solution; the exact modes, determined using the  spheroidal harmonics solution,  are shown in Fig.~\ref{fig:modes}a and excellent agreement is seen between the two.

		\begin{figure}[t!]
			\includegraphics[width=0.95\textwidth]{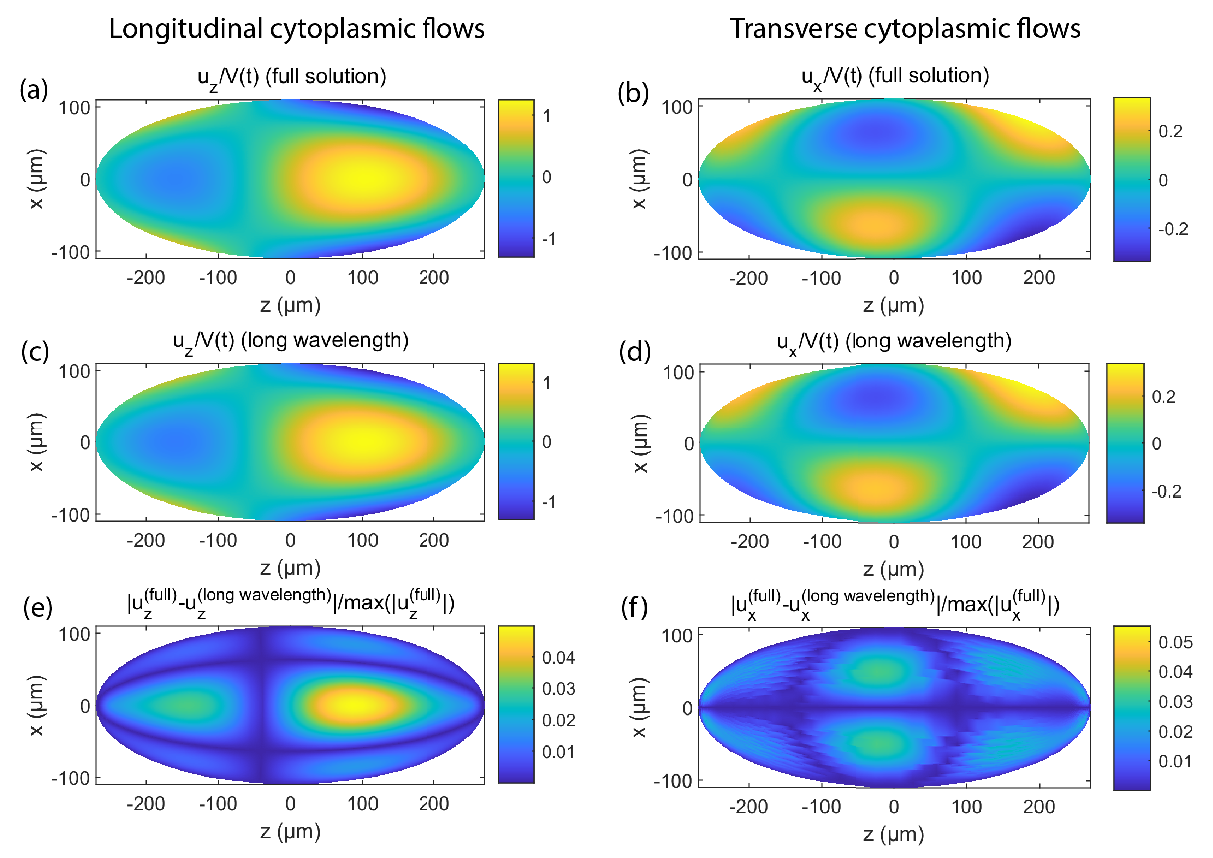}\caption{Cytoplasmic flows in the \textit{Drosophila} embryo using two different solution methods. Longitudinal component $u_z$ (a) and transverse component $u_x$ (b) of the cytoplasmic flow as produced by the full spheroidal harmonics solution. Longitudinal (c) and transverse (d) components  as produced by the long-wavelength solution. Errors incurred by the long-wavelength approximation for the longitudinal (e) and transverse (f) components of cytoplasmic flow, normalised by the maximum values of $u_z$ and $u_x$ respectively.}\label{fig:uv}
		\end{figure}

		We can then apply our long-wavelength solution to the cytoplasmic flows in a \textit{Drosophila} embryo i.e.~the flow driven by the slip velocity given by Eq.~\eqref{eq:modelflow}. In Fig.~\ref{fig:uv} we plot the longitudinal flows along the embryo axis, $u_z$, and the transverse flows, $u_x$, as determined by the full solution in spheroidal harmonics (Fig.~\ref{fig:uv}a-b), compared to the long-wavelength solution (Fig.~\ref{fig:uv}c-d). It is evident that the approximate solution is able to reproduce all features of the full solution. To quantify this further, we 
		display  the  errors  incurred by the long-wavelength approximation relative to the maximum velocity in the domain, $|u_i^{\text{full}}-u_i^{\text{long wavelength}}|/\max|u^{\text{full}}_i|$, in Fig.~\ref{fig:uv}e-f. Although the lubrication solution is strictly  only valid in the long-wavelength limit, we see that the long-wavelength solution reproduces a remarkably accurate approximation of the full solution in the entire embryo, incurring errors of less than $5\%$.

		\section{Reduced model of transport}
		\label{section:1D}
		
		In this final section, we show how to exploit the long-wavelength flow solution to derive a reduced-order model of nuclear transport. 
		By  projecting the full nuclear transport problem onto the embryo AP axis, we show that it admits an analytical solution that reproduces all the essential physics of the full system, including the characteristics of optimal transport.

		\subsection{Model formulation and solution}
		\label{sec:reducedmodel}
		We first define a number density function $n(z,t)$ of nuclei along the AP axis; this can be thought of as the projection of the density field along the central axis of the embryo. We then model the advection of nuclei by cytoplasmic flows as the advection of this   density function by the cytoplasmic velocity, $u_{AP}(z,t)$, along the AP axis. In the absence of diffusion, the field $n(z,t)$ is   governed by the one-dimensional advection equation
		\begin{equation}\label{eq:nori}
			\frac{\partial n}{\partial t} + \frac{\partial}{\partial z}(u_{AP}n)= 0.
		\end{equation}
		To further simplify the problem, we shift the asymmetric cortical flows, Eq.~\eqref{eq:modelflow}, into a symmetric sine profile, preserving its amplitude, and thus take
		\begin{equation}
			v_s(\zeta,t) = -V(t)\sin(\pi\zeta).
		\end{equation}
		This merely shifts the centre of the advective field but does not modify the essential features of the transport problem.

		From our lubrication solution, the cytoplasmic velocity $u_{AP}(z,t)$ along the AP axis may be expressed in terms of the cortical flow $v_s(\zeta,t)$ as
		\begin{equation}
			u_{AP}(z,t) = -v_s\left({z}/{b_z},t\right) \mathbf e_\zeta \cdot \mathbf e_z.
		\end{equation} 
		In the bulk,  $ \mathbf e_\zeta \cdot \mathbf e_z \approx 1$ because the cortex is approximately parallel to the AP axis; near the poles, where $\mathbf e_\zeta \cdot \mathbf e_z$ deviates significantly from unity,  $v_s$ is small and therefore the {absolute} difference between the component of $\mathbf v_s$ parallel to the cortex and the component parallel to the AP axis remains small.  Therefore, it is consistent to make the approximation 
		$u_{AP}(z,t) = -v_s\left({z}/{b_z},t\right) $ throughout.  Defining the nondimensionalised variable $Z := z/b_z$, we may therefore write
		\begin{equation}
			u_{AP}(Z,t) = V(t)\sin(\pi Z).
		\end{equation}
		Finally, we define a rescaled time
		\begin{equation}
			T := \frac{\pi}{b_z}\int_0^t{V(t')}\,\mathrm dt' = \frac{\pi}{b_z}\chi,
		\end{equation}
		so that, using these new variables, the transport equation,  Eq.~\eqref{eq:nori}, simplifies to 
		\begin{equation}\label{eq:1Dtransport}
			\frac{\partial n}{\partial T} + \frac{\partial}{\partial Z}\left[\frac{\sin(\pi\zeta)}{\pi}n\right] = 0. 
		\end{equation}

		Starting from the initial condition for the nuclei density $n(Z,0) := n_0(Z)$, we can use the method of characteristics to obtain the exact time-dependent solution as 
		\begin{equation}
			n(Z,T) = e^{-T}n_0\left( \frac{2}{\pi}\arctan\left(e^{-T}\tan\frac{\pi\zeta}{2}\right)\right)\frac{\tan^2\frac{\pi Z}{2} + 1}{e^{-2T}\tan^2\frac{\pi Z}{2} + 1}\cdot
		\end{equation}

		\subsection{Evolution of nuclear density and optimal axial spreading}
		
		In order to facilitate comparison between the full simulations (\S\ref{section:application}) and the reduced model (\S\ref{sec:reducedmodel}), we revert back to  the rescaled time variable, $\chi=Tb_z / \pi  $. We therefore now consider the density function $n(Z,\chi)$ as a function of the variables $Z$ and $\chi$.

		To show how the reduced model captures the essential features of the problem, we consider how a  cloud of nuclei centered at the origin is transported by it. We mimic the initial conditions used in the full simulations, i.e.~a uniform distribution of nuclei in a sphere of radius $60$ \textmu{m}, and thus make the specific choice  $n(Z,0):=n_0(Z) = \sqrt{\max(0,Z_0^2-Z^2)}/(\pi Z_0^2/2)$ with $Z_0 = $ (60~\textmu{m})$/b_z$.

		We plot in Fig.~\ref{fig:1D} the profiles of $n(Z,\chi)$  for  five values of the rescaled time $\chi$. 
		Similarly to what was seen in  the full simulations 
		(Fig.~\ref{fig:PHI}a, same values of $\chi$), the cytoplasmic flows initially spread the cloud of nuclei along the cell axis, but beyond some  value of $\chi$ applying further flows  hinders axial homogenisation and instead creates a `neck' scarce in nuclei in the centre of the cell.

		\begin{figure}[t!]
			\includegraphics[width=0.95\textwidth]{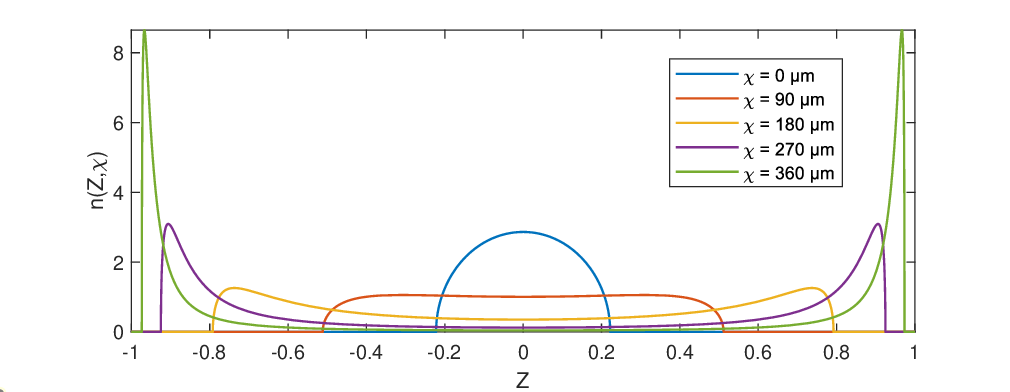}\caption{Nuclear transport by the reduced model. Profiles of $n(Z,\chi)$ for  increasing values of $\chi$, to be compared with Fig.~\ref{fig:PHI}a (same values of $\chi$). The variance of $n$, a measure of axial spreading, is plotted as a function of $\chi$ in Fig.~\ref{fig:PHI}b (purple dashed line).}\label{fig:1D}
		\end{figure}

		Similarly to our analysis of the full simulations, we may further quantify the nuclear spreading along the cell axis  by defining a measure of axial homogeneity for the reduced one-dimensional model. We compute the   variance,  $\Phi_{1D}(\chi)$, of $n(\chi,Z)$  over the full spatial interval $Z \in [-1,1]$:
		\begin{equation}
			\Phi_{1D}(\chi) := \int_{-1}^{1}[n(\chi,Z) - \bar n]^2\,\mathrm dZ,
		\end{equation}
		where $\bar n \equiv 1/2$ is the nuclear density averaged over $Z$  
		({note that} $\bar n$ is independent of $\chi$ since nuclei are conserved; this  may be seen directly  by  integrating Eq.~\eqref{eq:1Dtransport} over $Z$).		
		This variance is plotted against $\chi$ in Fig.~\ref{fig:PHI}b (purple dashed line). Remarkably, despite the drastic simplifications in the one-dimensional approach, the reduced model is able to capture all essential physics of nuclear transport, with axial spreading that is poor at low $\chi$ (nuclei initially concentrated in the centre of the cell) and high $\chi$ (nuclei accumulating at the poles; note that the 
		variance in the reduced  model overshoots that in the full simulations at large $\chi$ because  recirculation effects cannot be accounted for in one dimension).   
		Therefore, here again, axial nuclear transport is optimal for a finite value of $\chi$; the variance in the reduced model   is minimised at $\chi \approx 159$~\textmu{m} (purple diamond in Fig.~\ref{fig:PHI}b), which falls in between  the predicted optimal values $\chi \approx 195$~\textmu{m}, $213$~\textmu{m} obtained in the full model  with $R = 7$ \textmu{m}, $28$ \textmu{m} respectively (\S\ref{sec:optimal}), and the empirical value $\chi \approx 129$~\textmu{m} estimated from the  experiments (\S\ref{sec:comp}).

		\section{Discussion and conclusion}
		
		In this paper, motivated by cortex-driven flow-based transport of nuclei in the embryo of the model organism \textit{Drosophila melanogaster}, we  proposed two mathematical models to  study cytoplasmic streaming and boundary-driven flows inside elongated  biological cells. Assuming  (i) that the cytoplasm behaves like a Newtonian fluid at low Reynolds number, (ii) that the  cell  has a prolate spheroidal shape, and (iii) that the flow is driven by a prescribed axisymmetric tangential   velocity on the boundary of the cell, we are first able to compute the entire flow field analytically.

		To demonstrate the usefulness of  such an analytical modelling approach, we then considered     recent experiments characterising the transport of nuclei  in cell cycles 4-6 of  the   \textit{Drosophila} embryo.     By fitting the cortical contractions in our theoretical model to experimental data,  and introducing two different measures of axial spreading of the nuclei along the long axis of the embryo, we   reveal that experimental cortical flows ensure near-optimal
		spreading of the nuclei along the embryo.

		We next further simplify our theoretical approach and derive  a second model for the flow under the long-wavelength (lubrication) approximation. This approach provides a simpler solution for the flow -- and therefore a more practical one --  which reproduces the exact solution with  errors of $5\%$ or less without the need to evaluate  integrals of the boundary flow 	(see Eq.~\eqref{eq:bn}). We then used our long-wavelength solution in a reduced continuum model for nuclear transport, leading to  analytical solutions for the nuclear concentration that capture the essential physics of the full system, including optimal axial spreading.

		Obviously, the  modelling approach in this paper is focused on flow transport and 
		thus neglects some of the physico-chemical  parameters that may play key roles in cortical dynamics. { We have also modelled the system as a single Newtonian fluid driven by slip velocities at the boundary, and thus neglected possible complex rheological properties of the cytoplasm and cortex. This is in contrast to the recently introduced, more detailed   model  of the same system as an active actomyosin gel coupled to  a passive viscous cytoplasm~\cite{lopez2023}; this two-fluid model can be   solved numerically to quantitatively reproduce the detailed features of the experimental flow field, and in particular its deviations from a Stokes flow. However, a simple Stokes flow model as is used here does successfully explain the large-scale features of the flow and transport and} in turn, these  simplifications enable us to bypass the need for numerical computations in determining the flows and thus can be exploited to gain fundamental biophysical insight on the impact of cytoplasmic flows on cellular transport. The model is  versatile and may be adapted to the case of other cell shapes, or coupled with more complex models of the active cortex, for example,  one that includes  detailed biochemical feedback between cortical flows and nuclear positioning~\cite{lopez2023}.

				\section*{Acknowledgements}
		
		We thank Andrea Cairoli, Nikhil Desai and  Weida Liao for helpful comments on an earlier version of this manuscript. Financial support from the Cambridge Trust (scholarship to PHH) is gratefully acknowledged.

		\bibliography{drosophila1_references.bib}

\begin{thebibliography}{40}%
\makeatletter
\providecommand \@ifxundefined [1]{%
 \@ifx{#1\undefined}
}%
\providecommand \@ifnum [1]{%
 \ifnum #1\expandafter \@firstoftwo
 \else \expandafter \@secondoftwo
 \fi
}%
\providecommand \@ifx [1]{%
 \ifx #1\expandafter \@firstoftwo
 \else \expandafter \@secondoftwo
 \fi
}%
\providecommand \natexlab [1]{#1}%
\providecommand \enquote  [1]{``#1''}%
\providecommand \bibnamefont  [1]{#1}%
\providecommand \bibfnamefont [1]{#1}%
\providecommand \citenamefont [1]{#1}%
\providecommand \href@noop [0]{\@secondoftwo}%
\providecommand \href [0]{\begingroup \@sanitize@url \@href}%
\providecommand \@href[1]{\@@startlink{#1}\@@href}%
\providecommand \@@href[1]{\endgroup#1\@@endlink}%
\providecommand \@sanitize@url [0]{\catcode `\\12\catcode `\$12\catcode
  `\&12\catcode `\#12\catcode `\^12\catcode `\_12\catcode `\%12\relax}%
\providecommand \@@startlink[1]{}%
\providecommand \@@endlink[0]{}%
\providecommand \url  [0]{\begingroup\@sanitize@url \@url }%
\providecommand \@url [1]{\endgroup\@href {#1}{\urlprefix }}%
\providecommand \urlprefix  [0]{URL }%
\providecommand \Eprint [0]{\href }%
\providecommand \doibase [0]{https://doi.org/}%
\providecommand \selectlanguage [0]{\@gobble}%
\providecommand \bibinfo  [0]{\@secondoftwo}%
\providecommand \bibfield  [0]{\@secondoftwo}%
\providecommand \translation [1]{[#1]}%
\providecommand \BibitemOpen [0]{}%
\providecommand \bibitemStop [0]{}%
\providecommand \bibitemNoStop [0]{.\EOS\space}%
\providecommand \EOS [0]{\spacefactor3000\relax}%
\providecommand \BibitemShut  [1]{\csname bibitem#1\endcsname}%
\let\auto@bib@innerbib\@empty
%</preamble>
\bibitem [{\citenamefont {Salbreux}\ \emph {et~al.}(2012)\citenamefont
  {Salbreux}, \citenamefont {Charras},\ and\ \citenamefont
  {Paluch}}]{salbreux2012}%
  \BibitemOpen
  \bibfield  {author} {\bibinfo {author} {\bibfnamefont {G.}~\bibnamefont
  {Salbreux}}, \bibinfo {author} {\bibfnamefont {G.}~\bibnamefont {Charras}},\
  and\ \bibinfo {author} {\bibfnamefont {E.}~\bibnamefont {Paluch}},\
  }\bibfield  {title} {\bibinfo {title} {Actin cortex mechanics and cellular
  morphogenesis},\ }\href
  {https://doi.org/https://doi.org/10.1016/j.tcb.2012.07.001} {\bibfield
  {journal} {\bibinfo  {journal} {Trends Cell Biol.}\ }\textbf {\bibinfo
  {volume} {22}},\ \bibinfo {pages} {536} (\bibinfo {year} {2012})}\BibitemShut
  {NoStop}%
\bibitem [{\citenamefont {Chugh}\ and\ \citenamefont
  {Paluch}(2018)}]{chugh2018}%
  \BibitemOpen
  \bibfield  {author} {\bibinfo {author} {\bibfnamefont {P.}~\bibnamefont
  {Chugh}}\ and\ \bibinfo {author} {\bibfnamefont {E.~K.}\ \bibnamefont
  {Paluch}},\ }\bibfield  {title} {\bibinfo {title} {{The actin cortex at a
  glance}},\ }\href@noop {} {\bibfield  {journal} {\bibinfo  {journal} {J. Cell
  Sci.}\ }\textbf {\bibinfo {volume} {131}} (\bibinfo {year}
  {2018})}\BibitemShut {NoStop}%
\bibitem [{\citenamefont {Chabaud}\ \emph {et~al.}(2015)\citenamefont
  {Chabaud}, \citenamefont {Heuzé}, \citenamefont {Bretou}, \citenamefont
  {Vargas}, \citenamefont {Maiuri}, \citenamefont {Solanes}, \citenamefont
  {Maurin}, \citenamefont {Terriac}, \citenamefont {Le~Berre}, \citenamefont
  {Lankar},\ and\ \citenamefont {et~al.}}]{chabaud2015}%
  \BibitemOpen
  \bibfield  {author} {\bibinfo {author} {\bibfnamefont {M.}~\bibnamefont
  {Chabaud}}, \bibinfo {author} {\bibfnamefont {M.~L.}\ \bibnamefont {Heuzé}},
  \bibinfo {author} {\bibfnamefont {M.}~\bibnamefont {Bretou}}, \bibinfo
  {author} {\bibfnamefont {P.}~\bibnamefont {Vargas}}, \bibinfo {author}
  {\bibfnamefont {P.}~\bibnamefont {Maiuri}}, \bibinfo {author} {\bibfnamefont
  {P.}~\bibnamefont {Solanes}}, \bibinfo {author} {\bibfnamefont
  {M.}~\bibnamefont {Maurin}}, \bibinfo {author} {\bibfnamefont
  {E.}~\bibnamefont {Terriac}}, \bibinfo {author} {\bibfnamefont
  {M.}~\bibnamefont {Le~Berre}}, \bibinfo {author} {\bibfnamefont
  {D.}~\bibnamefont {Lankar}},\ and\ \bibinfo {author} {\bibnamefont
  {et~al.}},\ }\bibfield  {title} {\bibinfo {title} {Cell migration and antigen
  capture are antagonistic processes coupled by myosin ii in dendritic cells},\
  }\href@noop {} {\bibfield  {journal} {\bibinfo  {journal} {Nat. Comm.}\
  }\textbf {\bibinfo {volume} {6}} (\bibinfo {year} {2015})}\BibitemShut
  {NoStop}%
\bibitem [{\citenamefont {Bergert}\ \emph {et~al.}(2015)\citenamefont
  {Bergert}, \citenamefont {Erzberger}, \citenamefont {Desai}, \citenamefont
  {Aspalter}, \citenamefont {Oates}, \citenamefont {Charras}, \citenamefont
  {Salbreux},\ and\ \citenamefont {Paluch}}]{bergert2015}%
  \BibitemOpen
  \bibfield  {author} {\bibinfo {author} {\bibfnamefont {M.}~\bibnamefont
  {Bergert}}, \bibinfo {author} {\bibfnamefont {A.}~\bibnamefont {Erzberger}},
  \bibinfo {author} {\bibfnamefont {R.~A.}\ \bibnamefont {Desai}}, \bibinfo
  {author} {\bibfnamefont {I.~M.}\ \bibnamefont {Aspalter}}, \bibinfo {author}
  {\bibfnamefont {A.~C.}\ \bibnamefont {Oates}}, \bibinfo {author}
  {\bibfnamefont {G.}~\bibnamefont {Charras}}, \bibinfo {author} {\bibfnamefont
  {G.}~\bibnamefont {Salbreux}},\ and\ \bibinfo {author} {\bibfnamefont
  {E.~K.}\ \bibnamefont {Paluch}},\ }\bibfield  {title} {\bibinfo {title}
  {Force transmission during adhesion-independent migration},\ }\href
  {https://doi.org/10.1038/ncb3134} {\bibfield  {journal} {\bibinfo  {journal}
  {Nat. Cell Biol.}\ }\textbf {\bibinfo {volume} {17}},\ \bibinfo {pages}
  {524–529} (\bibinfo {year} {2015})}\BibitemShut {NoStop}%
\bibitem [{\citenamefont {Ruprecht}\ \emph {et~al.}(2015)\citenamefont
  {Ruprecht}, \citenamefont {Wieser}, \citenamefont {Callan-Jones},
  \citenamefont {Smutny}, \citenamefont {Morita}, \citenamefont {Sako},
  \citenamefont {Barone}, \citenamefont {Ritsch-Marte}, \citenamefont {Sixt},
  \citenamefont {Voituriez},\ and\ \citenamefont {Heisenberg}}]{ruprecht2015}%
  \BibitemOpen
  \bibfield  {author} {\bibinfo {author} {\bibfnamefont {V.}~\bibnamefont
  {Ruprecht}}, \bibinfo {author} {\bibfnamefont {S.}~\bibnamefont {Wieser}},
  \bibinfo {author} {\bibfnamefont {A.}~\bibnamefont {Callan-Jones}}, \bibinfo
  {author} {\bibfnamefont {M.}~\bibnamefont {Smutny}}, \bibinfo {author}
  {\bibfnamefont {H.}~\bibnamefont {Morita}}, \bibinfo {author} {\bibfnamefont
  {K.}~\bibnamefont {Sako}}, \bibinfo {author} {\bibfnamefont {V.}~\bibnamefont
  {Barone}}, \bibinfo {author} {\bibfnamefont {M.}~\bibnamefont
  {Ritsch-Marte}}, \bibinfo {author} {\bibfnamefont {M.}~\bibnamefont {Sixt}},
  \bibinfo {author} {\bibfnamefont {R.}~\bibnamefont {Voituriez}},\ and\
  \bibinfo {author} {\bibfnamefont {C.-P.}\ \bibnamefont {Heisenberg}},\
  }\bibfield  {title} {\bibinfo {title} {Cortical contractility triggers a
  stochastic switch to fast amoeboid cell motility},\ }\href
  {https://doi.org/https://doi.org/10.1016/j.cell.2015.01.008} {\bibfield
  {journal} {\bibinfo  {journal} {Cell}\ }\textbf {\bibinfo {volume} {160}},\
  \bibinfo {pages} {673} (\bibinfo {year} {2015})}\BibitemShut {NoStop}%
\bibitem [{\citenamefont {Liu}\ \emph {et~al.}(2015)\citenamefont {Liu},
  \citenamefont {Berre}, \citenamefont {Lautenschlaeger}, \citenamefont
  {Maiuri}, \citenamefont {Callan-Jones}, \citenamefont {Heuzé}, \citenamefont
  {Takaki}, \citenamefont {Voituriez},\ and\ \citenamefont {Piel}}]{liu2015}%
  \BibitemOpen
  \bibfield  {author} {\bibinfo {author} {\bibfnamefont {Y.-J.}\ \bibnamefont
  {Liu}}, \bibinfo {author} {\bibfnamefont {M.~L.}\ \bibnamefont {Berre}},
  \bibinfo {author} {\bibfnamefont {F.}~\bibnamefont {Lautenschlaeger}},
  \bibinfo {author} {\bibfnamefont {P.}~\bibnamefont {Maiuri}}, \bibinfo
  {author} {\bibfnamefont {A.}~\bibnamefont {Callan-Jones}}, \bibinfo {author}
  {\bibfnamefont {M.}~\bibnamefont {Heuzé}}, \bibinfo {author} {\bibfnamefont
  {T.}~\bibnamefont {Takaki}}, \bibinfo {author} {\bibfnamefont
  {R.}~\bibnamefont {Voituriez}},\ and\ \bibinfo {author} {\bibfnamefont
  {M.}~\bibnamefont {Piel}},\ }\bibfield  {title} {\bibinfo {title}
  {Confinement and low adhesion induce fast amoeboid migration of slow
  mesenchymal cells},\ }\href
  {https://doi.org/https://doi.org/10.1016/j.cell.2015.01.007} {\bibfield
  {journal} {\bibinfo  {journal} {Cell}\ }\textbf {\bibinfo {volume} {160}},\
  \bibinfo {pages} {659} (\bibinfo {year} {2015})}\BibitemShut {NoStop}%
\bibitem [{\citenamefont {Paluch}\ and\ \citenamefont
  {Raz}(2013)}]{paluch2013}%
  \BibitemOpen
  \bibfield  {author} {\bibinfo {author} {\bibfnamefont {E.~K.}\ \bibnamefont
  {Paluch}}\ and\ \bibinfo {author} {\bibfnamefont {E.}~\bibnamefont {Raz}},\
  }\bibfield  {title} {\bibinfo {title} {The role and regulation of blebs in
  cell migration},\ }\href
  {https://doi.org/https://doi.org/10.1016/j.ceb.2013.05.005} {\bibfield
  {journal} {\bibinfo  {journal} {Curr. Opin. Cell Biol.}\ }\textbf {\bibinfo
  {volume} {25}},\ \bibinfo {pages} {582} (\bibinfo {year} {2013})}\BibitemShut
  {NoStop}%
\bibitem [{\citenamefont {Blaser}\ \emph {et~al.}(2006)\citenamefont {Blaser},
  \citenamefont {Reichman-Fried}, \citenamefont {Castanon}, \citenamefont
  {Dumstrei}, \citenamefont {Marlow}, \citenamefont {Kawakami}, \citenamefont
  {Solnica-Krezel}, \citenamefont {Heisenberg},\ and\ \citenamefont
  {Raz}}]{blaser2006}%
  \BibitemOpen
  \bibfield  {author} {\bibinfo {author} {\bibfnamefont {H.}~\bibnamefont
  {Blaser}}, \bibinfo {author} {\bibfnamefont {M.}~\bibnamefont
  {Reichman-Fried}}, \bibinfo {author} {\bibfnamefont {I.}~\bibnamefont
  {Castanon}}, \bibinfo {author} {\bibfnamefont {K.}~\bibnamefont {Dumstrei}},
  \bibinfo {author} {\bibfnamefont {F.~L.}\ \bibnamefont {Marlow}}, \bibinfo
  {author} {\bibfnamefont {K.}~\bibnamefont {Kawakami}}, \bibinfo {author}
  {\bibfnamefont {L.}~\bibnamefont {Solnica-Krezel}}, \bibinfo {author}
  {\bibfnamefont {C.-P.}\ \bibnamefont {Heisenberg}},\ and\ \bibinfo {author}
  {\bibfnamefont {E.}~\bibnamefont {Raz}},\ }\bibfield  {title} {\bibinfo
  {title} {Migration of zebrafish primordial germ cells: A role for myosin
  contraction and cytoplasmic flow},\ }\href
  {https://doi.org/https://doi.org/10.1016/j.devcel.2006.09.023} {\bibfield
  {journal} {\bibinfo  {journal} {Dev. Cell}\ }\textbf {\bibinfo {volume}
  {11}},\ \bibinfo {pages} {613} (\bibinfo {year} {2006})}\BibitemShut
  {NoStop}%
\bibitem [{\citenamefont {Stewart}\ \emph {et~al.}(2011)\citenamefont
  {Stewart}, \citenamefont {Helenius}, \citenamefont {Toyoda}, \citenamefont
  {Ramanathan}, \citenamefont {Muller},\ and\ \citenamefont
  {Hyman}}]{stewart2011}%
  \BibitemOpen
  \bibfield  {author} {\bibinfo {author} {\bibfnamefont {M.~P.}\ \bibnamefont
  {Stewart}}, \bibinfo {author} {\bibfnamefont {J.}~\bibnamefont {Helenius}},
  \bibinfo {author} {\bibfnamefont {Y.}~\bibnamefont {Toyoda}}, \bibinfo
  {author} {\bibfnamefont {S.~P.}\ \bibnamefont {Ramanathan}}, \bibinfo
  {author} {\bibfnamefont {D.~J.}\ \bibnamefont {Muller}},\ and\ \bibinfo
  {author} {\bibfnamefont {A.~A.}\ \bibnamefont {Hyman}},\ }\bibfield  {title}
  {\bibinfo {title} {Hydrostatic pressure and the actomyosin cortex drive
  mitotic cell rounding},\ }\href {https://doi.org/10.1038/nature09642}
  {\bibfield  {journal} {\bibinfo  {journal} {Nature}\ }\textbf {\bibinfo
  {volume} {469}},\ \bibinfo {pages} {226–230} (\bibinfo {year}
  {2011})}\BibitemShut {NoStop}%
\bibitem [{\citenamefont {Schwayer}\ \emph {et~al.}(2016)\citenamefont
  {Schwayer}, \citenamefont {Sikora}, \citenamefont {Slováková},
  \citenamefont {Kardos},\ and\ \citenamefont {Heisenberg}}]{schwayer2016}%
  \BibitemOpen
  \bibfield  {author} {\bibinfo {author} {\bibfnamefont {C.}~\bibnamefont
  {Schwayer}}, \bibinfo {author} {\bibfnamefont {M.}~\bibnamefont {Sikora}},
  \bibinfo {author} {\bibfnamefont {J.}~\bibnamefont {Slováková}}, \bibinfo
  {author} {\bibfnamefont {R.}~\bibnamefont {Kardos}},\ and\ \bibinfo {author}
  {\bibfnamefont {C.-P.}\ \bibnamefont {Heisenberg}},\ }\bibfield  {title}
  {\bibinfo {title} {Actin rings of power},\ }\href
  {https://doi.org/https://doi.org/10.1016/j.devcel.2016.05.024} {\bibfield
  {journal} {\bibinfo  {journal} {Dev. Cell}\ }\textbf {\bibinfo {volume}
  {37}},\ \bibinfo {pages} {493} (\bibinfo {year} {2016})}\BibitemShut
  {NoStop}%
\bibitem [{\citenamefont {Lu}\ and\ \citenamefont {Gelfand}(2022)}]{lu2022}%
  \BibitemOpen
  \bibfield  {author} {\bibinfo {author} {\bibfnamefont {W.}~\bibnamefont
  {Lu}}\ and\ \bibinfo {author} {\bibfnamefont {V.~I.}\ \bibnamefont
  {Gelfand}},\ }\bibfield  {title} {\bibinfo {title} {{Go with the flow –
  bulk transport by molecular motors}},\ }\href@noop {} {\bibfield  {journal}
  {\bibinfo  {journal} {J. Cell Sci.}\ }\textbf {\bibinfo {volume} {136}}
  (\bibinfo {year} {2022})}\BibitemShut {NoStop}%
\bibitem [{\citenamefont {Goldstein}\ and\ \citenamefont {van~de
  Meent}(2015)}]{goldstein2015}%
  \BibitemOpen
  \bibfield  {author} {\bibinfo {author} {\bibfnamefont {R.~E.}\ \bibnamefont
  {Goldstein}}\ and\ \bibinfo {author} {\bibfnamefont {J.-W.}\ \bibnamefont
  {van~de Meent}},\ }\bibfield  {title} {\bibinfo {title} {A physical
  perspective on cytoplasmic streaming},\ }\href
  {https://doi.org/10.1098/rsfs.2015.0030} {\bibfield  {journal} {\bibinfo
  {journal} {Interface Focus}\ }\textbf {\bibinfo {volume} {5}},\ \bibinfo
  {pages} {20150030} (\bibinfo {year} {2015})}\BibitemShut {NoStop}%
\bibitem [{\citenamefont {Illukkumbura}\ \emph {et~al.}(2020)\citenamefont
  {Illukkumbura}, \citenamefont {Bland},\ and\ \citenamefont
  {Goehring}}]{illukkumbura2020}%
  \BibitemOpen
  \bibfield  {author} {\bibinfo {author} {\bibfnamefont {R.}~\bibnamefont
  {Illukkumbura}}, \bibinfo {author} {\bibfnamefont {T.}~\bibnamefont
  {Bland}},\ and\ \bibinfo {author} {\bibfnamefont {N.~W.}\ \bibnamefont
  {Goehring}},\ }\bibfield  {title} {\bibinfo {title} {Patterning and
  polarization of cells by intracellular flows},\ }\href
  {https://doi.org/https://doi.org/10.1016/j.ceb.2019.10.005} {\bibfield
  {journal} {\bibinfo  {journal} {Current Opinion in Cell Biology}\ }\textbf
  {\bibinfo {volume} {62}},\ \bibinfo {pages} {123} (\bibinfo {year} {2020})},\
  \bibinfo {note} {cell Architecture}\BibitemShut {NoStop}%
\bibitem [{\citenamefont {Pieuchot}\ \emph {et~al.}(2015)\citenamefont
  {Pieuchot}, \citenamefont {Lai}, \citenamefont {Loh}, \citenamefont {Leong},
  \citenamefont {Chiam}, \citenamefont {Stajich},\ and\ \citenamefont
  {Jedd}}]{pieuchot2015}%
  \BibitemOpen
  \bibfield  {author} {\bibinfo {author} {\bibfnamefont {L.}~\bibnamefont
  {Pieuchot}}, \bibinfo {author} {\bibfnamefont {J.}~\bibnamefont {Lai}},
  \bibinfo {author} {\bibfnamefont {R.~A.}\ \bibnamefont {Loh}}, \bibinfo
  {author} {\bibfnamefont {F.~Y.}\ \bibnamefont {Leong}}, \bibinfo {author}
  {\bibfnamefont {K.-H.}\ \bibnamefont {Chiam}}, \bibinfo {author}
  {\bibfnamefont {J.}~\bibnamefont {Stajich}},\ and\ \bibinfo {author}
  {\bibfnamefont {G.}~\bibnamefont {Jedd}},\ }\bibfield  {title} {\bibinfo
  {title} {Cellular subcompartments through cytoplasmic streaming},\ }\href
  {https://doi.org/https://doi.org/10.1016/j.devcel.2015.07.017} {\bibfield
  {journal} {\bibinfo  {journal} {Developmental Cell}\ }\textbf {\bibinfo
  {volume} {34}},\ \bibinfo {pages} {410} (\bibinfo {year} {2015})}\BibitemShut
  {NoStop}%
\bibitem [{\citenamefont {Fuentes}\ and\ \citenamefont
  {Fernández}(2010)}]{fuentes2010}%
  \BibitemOpen
  \bibfield  {author} {\bibinfo {author} {\bibfnamefont {R.}~\bibnamefont
  {Fuentes}}\ and\ \bibinfo {author} {\bibfnamefont {J.}~\bibnamefont
  {Fernández}},\ }\bibfield  {title} {\bibinfo {title} {Ooplasmic segregation
  in the zebrafish zygote and early embryo: Pattern of ooplasmic movements and
  transport pathways},\ }\href
  {https://doi.org/https://doi.org/10.1002/dvdy.22349} {\bibfield  {journal}
  {\bibinfo  {journal} {Dev. Dyn.}\ }\textbf {\bibinfo {volume} {239}},\
  \bibinfo {pages} {2172} (\bibinfo {year} {2010})}\BibitemShut {NoStop}%
\bibitem [{\citenamefont {Shamipour}\ \emph {et~al.}(2019)\citenamefont
  {Shamipour}, \citenamefont {Kardos}, \citenamefont {Xue}, \citenamefont
  {Hof}, \citenamefont {Hannezo},\ and\ \citenamefont
  {Heisenberg}}]{shamipour2019}%
  \BibitemOpen
  \bibfield  {author} {\bibinfo {author} {\bibfnamefont {S.}~\bibnamefont
  {Shamipour}}, \bibinfo {author} {\bibfnamefont {R.}~\bibnamefont {Kardos}},
  \bibinfo {author} {\bibfnamefont {S.-L.}\ \bibnamefont {Xue}}, \bibinfo
  {author} {\bibfnamefont {B.}~\bibnamefont {Hof}}, \bibinfo {author}
  {\bibfnamefont {E.}~\bibnamefont {Hannezo}},\ and\ \bibinfo {author}
  {\bibfnamefont {C.-P.}\ \bibnamefont {Heisenberg}},\ }\bibfield  {title}
  {\bibinfo {title} {Bulk actin dynamics drive phase segregation in zebrafish
  oocytes},\ }\href
  {https://doi.org/https://doi.org/10.1016/j.cell.2019.04.030} {\bibfield
  {journal} {\bibinfo  {journal} {Cell}\ }\textbf {\bibinfo {volume} {177}},\
  \bibinfo {pages} {1463} (\bibinfo {year} {2019})}\BibitemShut {NoStop}%
\bibitem [{\citenamefont {Gubieda}\ \emph {et~al.}(2020)\citenamefont
  {Gubieda}, \citenamefont {Packer}, \citenamefont {Squires}, \citenamefont
  {Martin},\ and\ \citenamefont {Rodriguez}}]{gubieda2020}%
  \BibitemOpen
  \bibfield  {author} {\bibinfo {author} {\bibfnamefont {A.~G.}\ \bibnamefont
  {Gubieda}}, \bibinfo {author} {\bibfnamefont {J.~R.}\ \bibnamefont {Packer}},
  \bibinfo {author} {\bibfnamefont {I.}~\bibnamefont {Squires}}, \bibinfo
  {author} {\bibfnamefont {J.}~\bibnamefont {Martin}},\ and\ \bibinfo {author}
  {\bibfnamefont {J.}~\bibnamefont {Rodriguez}},\ }\bibfield  {title} {\bibinfo
  {title} {Going with the flow: insights from \textit{Caenorhabditis elegans}
  zygote polarization},\ }\href {https://doi.org/10.1098/rstb.2019.0555}
  {\bibfield  {journal} {\bibinfo  {journal} {Philos. Trans. R. Soc. B: Biol.
  Sci.}\ }\textbf {\bibinfo {volume} {375}},\ \bibinfo {pages} {20190555}
  (\bibinfo {year} {2020})}\BibitemShut {NoStop}%
\bibitem [{\citenamefont {Deneke}\ \emph {et~al.}(2019)\citenamefont {Deneke},
  \citenamefont {Puliafito}, \citenamefont {Krueger}, \citenamefont {Narla},
  \citenamefont {{De Simone}}, \citenamefont {Primo}, \citenamefont
  {Vergassola}, \citenamefont {{De Renzis}},\ and\ \citenamefont {{Di
  Talia}}}]{deneke2019}%
  \BibitemOpen
  \bibfield  {author} {\bibinfo {author} {\bibfnamefont {V.~E.}\ \bibnamefont
  {Deneke}}, \bibinfo {author} {\bibfnamefont {A.}~\bibnamefont {Puliafito}},
  \bibinfo {author} {\bibfnamefont {D.}~\bibnamefont {Krueger}}, \bibinfo
  {author} {\bibfnamefont {A.~V.}\ \bibnamefont {Narla}}, \bibinfo {author}
  {\bibfnamefont {A.}~\bibnamefont {{De Simone}}}, \bibinfo {author}
  {\bibfnamefont {L.}~\bibnamefont {Primo}}, \bibinfo {author} {\bibfnamefont
  {M.}~\bibnamefont {Vergassola}}, \bibinfo {author} {\bibfnamefont
  {S.}~\bibnamefont {{De Renzis}}},\ and\ \bibinfo {author} {\bibfnamefont
  {S.}~\bibnamefont {{Di Talia}}},\ }\bibfield  {title} {\bibinfo {title}
  {Self-organized nuclear positioning synchronizes the cell cycle in
  \textit{Drosophila} embryos},\ }\href
  {https://doi.org/https://doi.org/10.1016/j.cell.2019.03.007} {\bibfield
  {journal} {\bibinfo  {journal} {Cell}\ }\textbf {\bibinfo {volume} {177}},\
  \bibinfo {pages} {925} (\bibinfo {year} {2019})}\BibitemShut {NoStop}%
\bibitem [{\citenamefont {Prost}\ \emph {et~al.}(2015)\citenamefont {Prost},
  \citenamefont {Jülicher},\ and\ \citenamefont {Joanny}}]{prost2015}%
  \BibitemOpen
  \bibfield  {author} {\bibinfo {author} {\bibfnamefont {J.}~\bibnamefont
  {Prost}}, \bibinfo {author} {\bibfnamefont {F.}~\bibnamefont {Jülicher}},\
  and\ \bibinfo {author} {\bibfnamefont {J.-F.}\ \bibnamefont {Joanny}},\
  }\bibfield  {title} {\bibinfo {title} {Active gel physics},\ }\href
  {https://doi.org/10.1038/nphys3224} {\bibfield  {journal} {\bibinfo
  {journal} {Nat. Phys.}\ }\textbf {\bibinfo {volume} {11}},\ \bibinfo {pages}
  {111–117} (\bibinfo {year} {2015})}\BibitemShut {NoStop}%
\bibitem [{\citenamefont {Salbreux}\ and\ \citenamefont
  {J\"ulicher}(2017)}]{salbreux2017}%
  \BibitemOpen
  \bibfield  {author} {\bibinfo {author} {\bibfnamefont {G.}~\bibnamefont
  {Salbreux}}\ and\ \bibinfo {author} {\bibfnamefont {F.}~\bibnamefont
  {J\"ulicher}},\ }\bibfield  {title} {\bibinfo {title} {Mechanics of active
  surfaces},\ }\href {https://doi.org/10.1103/PhysRevE.96.032404} {\bibfield
  {journal} {\bibinfo  {journal} {Phys. Rev. E}\ }\textbf {\bibinfo {volume}
  {96}},\ \bibinfo {pages} {032404} (\bibinfo {year} {2017})}\BibitemShut
  {NoStop}%
\bibitem [{\citenamefont {{Borja da Rocha}}\ \emph {et~al.}(2022)\citenamefont
  {{Borja da Rocha}}, \citenamefont {Bleyer},\ and\ \citenamefont
  {Turlier}}]{rocha2022}%
  \BibitemOpen
  \bibfield  {author} {\bibinfo {author} {\bibfnamefont {H.}~\bibnamefont
  {{Borja da Rocha}}}, \bibinfo {author} {\bibfnamefont {J.}~\bibnamefont
  {Bleyer}},\ and\ \bibinfo {author} {\bibfnamefont {H.}~\bibnamefont
  {Turlier}},\ }\bibfield  {title} {\bibinfo {title} {A viscous active shell
  theory of the cell cortex},\ }\href
  {https://doi.org/https://doi.org/10.1016/j.jmps.2022.104876} {\bibfield
  {journal} {\bibinfo  {journal} {J. Mech. Phys. Solids.}\ }\textbf {\bibinfo
  {volume} {164}},\ \bibinfo {pages} {104876} (\bibinfo {year}
  {2022})}\BibitemShut {NoStop}%
\bibitem [{\citenamefont {Torres-Sánchez}\ \emph {et~al.}(2019)\citenamefont
  {Torres-Sánchez}, \citenamefont {Millán},\ and\ \citenamefont
  {Arroyo}}]{torressanchez2019}%
  \BibitemOpen
  \bibfield  {author} {\bibinfo {author} {\bibfnamefont {A.}~\bibnamefont
  {Torres-Sánchez}}, \bibinfo {author} {\bibfnamefont {D.}~\bibnamefont
  {Millán}},\ and\ \bibinfo {author} {\bibfnamefont {M.}~\bibnamefont
  {Arroyo}},\ }\bibfield  {title} {\bibinfo {title} {Modelling fluid deformable
  surfaces with an emphasis on biological interfaces},\ }\href
  {https://doi.org/10.1017/jfm.2019.341} {\bibfield  {journal} {\bibinfo
  {journal} {J. Fluid Mech.}\ }\textbf {\bibinfo {volume} {872}},\ \bibinfo
  {pages} {218–271} (\bibinfo {year} {2019})}\BibitemShut {NoStop}%
\bibitem [{\citenamefont {Bächer}\ \emph {et~al.}(2021)\citenamefont
  {Bächer}, \citenamefont {Khoromskaia}, \citenamefont {Salbreux},\ and\
  \citenamefont {Gekle}}]{bacher2021}%
  \BibitemOpen
  \bibfield  {author} {\bibinfo {author} {\bibfnamefont {C.}~\bibnamefont
  {Bächer}}, \bibinfo {author} {\bibfnamefont {D.}~\bibnamefont
  {Khoromskaia}}, \bibinfo {author} {\bibfnamefont {G.}~\bibnamefont
  {Salbreux}},\ and\ \bibinfo {author} {\bibfnamefont {S.}~\bibnamefont
  {Gekle}},\ }\bibfield  {title} {\bibinfo {title} {A three-dimensional
  numerical model of an active cell cortex in the viscous limit},\ }\href@noop
  {} {\bibfield  {journal} {\bibinfo  {journal} {Front. Phys.}\ }\textbf
  {\bibinfo {volume} {9}} (\bibinfo {year} {2021})}\BibitemShut {NoStop}%
\bibitem [{\citenamefont {B\"acher}\ and\ \citenamefont
  {Gekle}(2019)}]{bacher2019}%
  \BibitemOpen
  \bibfield  {author} {\bibinfo {author} {\bibfnamefont {C.}~\bibnamefont
  {B\"acher}}\ and\ \bibinfo {author} {\bibfnamefont {S.}~\bibnamefont
  {Gekle}},\ }\bibfield  {title} {\bibinfo {title} {Computational modeling of
  active deformable membranes embedded in three-dimensional flows},\ }\href
  {https://doi.org/10.1103/PhysRevE.99.062418} {\bibfield  {journal} {\bibinfo
  {journal} {Phys. Rev. E}\ }\textbf {\bibinfo {volume} {99}},\ \bibinfo
  {pages} {062418} (\bibinfo {year} {2019})}\BibitemShut {NoStop}%
\bibitem [{\citenamefont {L.}\ \emph {et~al.}(2023)\citenamefont {L.},
  \citenamefont {Puliafito}, \citenamefont {Xu}, \citenamefont {Lu},
  \citenamefont {{Di Talia}},\ and\ \citenamefont {Vergassola}}]{lopez2023}%
  \BibitemOpen
  \bibfield  {author} {\bibinfo {author} {\bibfnamefont {C.~H.}\ \bibnamefont
  {L.}}, \bibinfo {author} {\bibfnamefont {A.}~\bibnamefont {Puliafito}},
  \bibinfo {author} {\bibfnamefont {Y.}~\bibnamefont {Xu}}, \bibinfo {author}
  {\bibfnamefont {Z.}~\bibnamefont {Lu}}, \bibinfo {author} {\bibfnamefont
  {S.}~\bibnamefont {{Di Talia}}},\ and\ \bibinfo {author} {\bibfnamefont
  {M.}~\bibnamefont {Vergassola}},\ }\bibfield  {title} {\bibinfo {title}
  {Two-fluid dynamics and micron-thin boundary layers shape cytoplasmic flows
  in early \textit{Drosophila} embryos}} (\bibinfo {year} {2023})\BibitemShut
  {NoStop}%
\bibitem [{\citenamefont {Klughammer}\ \emph {et~al.}(2018)\citenamefont
  {Klughammer}, \citenamefont {Bischof}, \citenamefont {Schnellbächer},
  \citenamefont {Callegari}, \citenamefont {Lénárt},\ and\ \citenamefont
  {Schwarz}}]{klughammer2018}%
  \BibitemOpen
  \bibfield  {author} {\bibinfo {author} {\bibfnamefont {N.}~\bibnamefont
  {Klughammer}}, \bibinfo {author} {\bibfnamefont {J.}~\bibnamefont {Bischof}},
  \bibinfo {author} {\bibfnamefont {N.~D.}\ \bibnamefont {Schnellbächer}},
  \bibinfo {author} {\bibfnamefont {A.}~\bibnamefont {Callegari}}, \bibinfo
  {author} {\bibfnamefont {P.}~\bibnamefont {Lénárt}},\ and\ \bibinfo
  {author} {\bibfnamefont {U.~S.}\ \bibnamefont {Schwarz}},\ }\bibfield
  {title} {\bibinfo {title} {Cytoplasmic flows in starfish oocytes are fully
  determined by cortical contractions},\ }\href
  {https://doi.org/10.1371/journal.pcbi.1006588} {\bibfield  {journal}
  {\bibinfo  {journal} {PLoS Comput. Biol.}\ }\textbf {\bibinfo {volume}
  {14}},\ \bibinfo {pages} {1} (\bibinfo {year} {2018})}\BibitemShut {NoStop}%
\bibitem [{\citenamefont {Niwayama}\ \emph {et~al.}(2011)\citenamefont
  {Niwayama}, \citenamefont {Shinohara},\ and\ \citenamefont
  {Kimura}}]{niwayama2011}%
  \BibitemOpen
  \bibfield  {author} {\bibinfo {author} {\bibfnamefont {R.}~\bibnamefont
  {Niwayama}}, \bibinfo {author} {\bibfnamefont {K.}~\bibnamefont
  {Shinohara}},\ and\ \bibinfo {author} {\bibfnamefont {A.}~\bibnamefont
  {Kimura}},\ }\bibfield  {title} {\bibinfo {title} {Hydrodynamic property of
  the cytoplasm is sufficient to mediate cytoplasmic streaming in the
  \textit{Caenorhabiditis elegans} embryo},\ }\href
  {https://doi.org/10.1073/pnas.1101853108} {\bibfield  {journal} {\bibinfo
  {journal} {Proceedings of the National Academy of Sciences}\ }\textbf
  {\bibinfo {volume} {108}},\ \bibinfo {pages} {11900} (\bibinfo {year}
  {2011})}\BibitemShut {NoStop}%
\bibitem [{\citenamefont {Ganguly}\ \emph {et~al.}(2012)\citenamefont
  {Ganguly}, \citenamefont {Williams}, \citenamefont {Palacios},\ and\
  \citenamefont {Goldstein}}]{ganguly2012}%
  \BibitemOpen
  \bibfield  {author} {\bibinfo {author} {\bibfnamefont {S.}~\bibnamefont
  {Ganguly}}, \bibinfo {author} {\bibfnamefont {L.~S.}\ \bibnamefont
  {Williams}}, \bibinfo {author} {\bibfnamefont {I.~M.}\ \bibnamefont
  {Palacios}},\ and\ \bibinfo {author} {\bibfnamefont {R.~E.}\ \bibnamefont
  {Goldstein}},\ }\bibfield  {title} {\bibinfo {title} {Cytoplasmic streaming
  in \textit{Drosophila} oocytes varies with kinesin activity and correlates
  with the microtubule cytoskeleton architecture},\ }\href
  {https://doi.org/10.1073/pnas.1203575109} {\bibfield  {journal} {\bibinfo
  {journal} {Proceedings of the National Academy of Sciences}\ }\textbf
  {\bibinfo {volume} {109}},\ \bibinfo {pages} {15109} (\bibinfo {year}
  {2012})}\BibitemShut {NoStop}%
\bibitem [{\citenamefont {Happel}\ and\ \citenamefont
  {Brenner}(1965)}]{happel1965}%
  \BibitemOpen
  \bibfield  {author} {\bibinfo {author} {\bibfnamefont {J.}~\bibnamefont
  {Happel}}\ and\ \bibinfo {author} {\bibfnamefont {H.}~\bibnamefont
  {Brenner}},\ }\href@noop {} {\emph {\bibinfo {title} {Low Reynolds number
  hydrodynamics}}}\ (\bibinfo  {publisher} {Prentice-Hall},\ \bibinfo {year}
  {1965})\BibitemShut {NoStop}%
\bibitem [{\citenamefont {Pöhnl}\ \emph {et~al.}(2020)\citenamefont {Pöhnl},
  \citenamefont {Popescu},\ and\ \citenamefont {Uspal}}]{pohnl2020}%
  \BibitemOpen
  \bibfield  {author} {\bibinfo {author} {\bibfnamefont {R.}~\bibnamefont
  {Pöhnl}}, \bibinfo {author} {\bibfnamefont {M.~N.}\ \bibnamefont
  {Popescu}},\ and\ \bibinfo {author} {\bibfnamefont {W.~E.}\ \bibnamefont
  {Uspal}},\ }\bibfield  {title} {\bibinfo {title} {Axisymmetric spheroidal
  squirmers and self-diffusiophoretic particles},\ }\href
  {https://doi.org/10.1088/1361-648X/ab5edd} {\bibfield  {journal} {\bibinfo
  {journal} {J. Phys. Condens. Matter}\ }\textbf {\bibinfo {volume} {32}},\
  \bibinfo {pages} {164001} (\bibinfo {year} {2020})}\BibitemShut {NoStop}%
\bibitem [{\citenamefont {Dassios}\ \emph {et~al.}(1994)\citenamefont
  {Dassios}, \citenamefont {Hadjinicolaou},\ and\ \citenamefont
  {Payatakes}}]{dassios1994}%
  \BibitemOpen
  \bibfield  {author} {\bibinfo {author} {\bibfnamefont {G.}~\bibnamefont
  {Dassios}}, \bibinfo {author} {\bibfnamefont {M.}~\bibnamefont
  {Hadjinicolaou}},\ and\ \bibinfo {author} {\bibfnamefont {A.~C.}\
  \bibnamefont {Payatakes}},\ }\href@noop {} {\bibfield  {journal} {\bibinfo
  {journal} {Q. Appl. Math.}\ }\textbf {\bibinfo {volume} {52}},\ \bibinfo
  {pages} {157} (\bibinfo {year} {1994})}\BibitemShut {NoStop}%
\bibitem [{\citenamefont {Abramowitz}\ and\ \citenamefont
  {Stegun}(2013)}]{abramowitz2013}%
  \BibitemOpen
  \bibfield  {author} {\bibinfo {author} {\bibfnamefont {M.}~\bibnamefont
  {Abramowitz}}\ and\ \bibinfo {author} {\bibfnamefont {I.~A.}\ \bibnamefont
  {Stegun}},\ }\href@noop {} {\emph {\bibinfo {title} {Handbook of Mathematical
  Functions with formulas, graphs, and mathematical tables}}}\ (\bibinfo
  {publisher} {Dover Publ},\ \bibinfo {year} {2013})\BibitemShut {NoStop}%
\bibitem [{\citenamefont {Farrell}\ and\ \citenamefont
  {O'Farrell}(2014)}]{farell2014}%
  \BibitemOpen
  \bibfield  {author} {\bibinfo {author} {\bibfnamefont {J.~A.}\ \bibnamefont
  {Farrell}}\ and\ \bibinfo {author} {\bibfnamefont {P.~H.}\ \bibnamefont
  {O'Farrell}},\ }\bibfield  {title} {\bibinfo {title} {From egg to gastrula:
  How the cell cycle is remodeled during the \textit{Drosophila} mid-blastula
  transition},\ }\href {https://doi.org/10.1146/annurev-genet-111212-133531}
  {\bibfield  {journal} {\bibinfo  {journal} {Annu. Rev. Genet.}\ }\textbf
  {\bibinfo {volume} {48}},\ \bibinfo {pages} {269} (\bibinfo {year}
  {2014})}\BibitemShut {NoStop}%
\bibitem [{\citenamefont {Rabinowitz}(1941)}]{rabinowitz1941}%
  \BibitemOpen
  \bibfield  {author} {\bibinfo {author} {\bibfnamefont {M.}~\bibnamefont
  {Rabinowitz}},\ }\bibfield  {title} {\bibinfo {title} {Studies on the
  cytology and early embryology of the egg of \textit{Drosophila
  melanogaster}},\ }\href
  {https://doi.org/https://doi.org/10.1002/jmor.1050690102} {\bibfield
  {journal} {\bibinfo  {journal} {J. Morphol.}\ }\textbf {\bibinfo {volume}
  {69}},\ \bibinfo {pages} {1} (\bibinfo {year} {1941})}\BibitemShut {NoStop}%
\bibitem [{\citenamefont {Zalokar}\ and\ \citenamefont
  {Erk}(1976)}]{zalokar1976}%
  \BibitemOpen
  \bibfield  {author} {\bibinfo {author} {\bibfnamefont {M.}~\bibnamefont
  {Zalokar}}\ and\ \bibinfo {author} {\bibfnamefont {I.}~\bibnamefont {Erk}},\
  }\bibfield  {title} {\bibinfo {title} {Division and migration of nuclei
  during early embryogenesis of \textit{Drosophila melanogaster}},\ }\href@noop
  {} {\bibfield  {journal} {\bibinfo  {journal} {J. Microsc. Biol. Cell.}\
  }\textbf {\bibinfo {volume} {25}},\ \bibinfo {pages} {97 – 106} (\bibinfo
  {year} {1976})}\BibitemShut {NoStop}%
\bibitem [{\citenamefont {Swaminathan}\ \emph {et~al.}(1997)\citenamefont
  {Swaminathan}, \citenamefont {Hoang},\ and\ \citenamefont
  {Verkman}}]{swaminathan1997}%
  \BibitemOpen
  \bibfield  {author} {\bibinfo {author} {\bibfnamefont {R.}~\bibnamefont
  {Swaminathan}}, \bibinfo {author} {\bibfnamefont {C.}~\bibnamefont {Hoang}},\
  and\ \bibinfo {author} {\bibfnamefont {A.}~\bibnamefont {Verkman}},\
  }\bibfield  {title} {\bibinfo {title} {Photobleaching recovery and anisotropy
  decay of green fluorescent protein gfp-s65t in solution and cells:
  cytoplasmic viscosity probed by green fluorescent protein translational and
  rotational diffusion},\ }\href
  {https://doi.org/https://doi.org/10.1016/S0006-3495(97)78835-0} {\bibfield
  {journal} {\bibinfo  {journal} {Biophys. J.}\ }\textbf {\bibinfo {volume}
  {72}},\ \bibinfo {pages} {1900} (\bibinfo {year} {1997})}\BibitemShut
  {NoStop}%
\bibitem [{\citenamefont {Kim}\ and\ \citenamefont {Karrila}(1991)}]{kimbook}%
  \BibitemOpen
  \bibfield  {author} {\bibinfo {author} {\bibfnamefont {S.}~\bibnamefont
  {Kim}}\ and\ \bibinfo {author} {\bibfnamefont {J.~S.}\ \bibnamefont
  {Karrila}},\ }\href@noop {} {\emph {\bibinfo {title} {Microhydrodynamics:
  {P}rinciples and {S}elected {A}pplications.}}}\ (\bibinfo  {publisher}
  {Butterworth-Heinemann},\ \bibinfo {address} {Boston, MA},\ \bibinfo {year}
  {1991})\BibitemShut {NoStop}%
\bibitem [{\citenamefont {Sullivan}\ and\ \citenamefont
  {Theurkauf}(1995)}]{sullivan1995}%
  \BibitemOpen
  \bibfield  {author} {\bibinfo {author} {\bibfnamefont {W.}~\bibnamefont
  {Sullivan}}\ and\ \bibinfo {author} {\bibfnamefont {W.~E.}\ \bibnamefont
  {Theurkauf}},\ }\bibfield  {title} {\bibinfo {title} {The cytoskeleton and
  morphogenesis of the early \textit{Drosophila} embryo},\ }\href
  {https://doi.org/https://doi.org/10.1016/0955-0674(95)80040-9} {\bibfield
  {journal} {\bibinfo  {journal} {Curr. Opin. Cell Biol.}\ }\textbf {\bibinfo
  {volume} {7}},\ \bibinfo {pages} {18} (\bibinfo {year} {1995})}\BibitemShut
  {NoStop}%
\bibitem [{\citenamefont {Telley}\ \emph {et~al.}(2012)\citenamefont {Telley},
  \citenamefont {Gáspár}, \citenamefont {Ephrussi},\ and\ \citenamefont
  {Surrey}}]{telley2012}%
  \BibitemOpen
  \bibfield  {author} {\bibinfo {author} {\bibfnamefont {I.~A.}\ \bibnamefont
  {Telley}}, \bibinfo {author} {\bibfnamefont {I.}~\bibnamefont {Gáspár}},
  \bibinfo {author} {\bibfnamefont {A.}~\bibnamefont {Ephrussi}},\ and\
  \bibinfo {author} {\bibfnamefont {T.}~\bibnamefont {Surrey}},\ }\bibfield
  {title} {\bibinfo {title} {{Aster migration determines the length scale of
  nuclear separation in the Drosophila syncytial embryo}},\ }\href
  {https://doi.org/10.1083/jcb.201204019} {\bibfield  {journal} {\bibinfo
  {journal} {Journal of Cell Biology}\ }\textbf {\bibinfo {volume} {197}},\
  \bibinfo {pages} {887} (\bibinfo {year} {2012})}\BibitemShut {NoStop}%
\bibitem [{\citenamefont {Leal}(2007)}]{leal2007}%
  \BibitemOpen
  \bibfield  {author} {\bibinfo {author} {\bibfnamefont {L.~G.}\ \bibnamefont
  {Leal}},\ }\href {https://doi.org/10.1017/CBO9780511800245} {\emph {\bibinfo
  {title} {Advanced Transport Phenomena: Fluid Mechanics and Convective
  Transport Processes}}},\ Cambridge Series in Chemical Engineering\ (\bibinfo
  {publisher} {Cambridge University Press},\ \bibinfo {year}
  {2007})\BibitemShut {NoStop}%
\end{thebibliography}%
		
	\end{document}